\documentclass[traditabstract]{aa}
\usepackage{graphicx}
\usepackage{longtable}
\usepackage{footnote}
\usepackage{xcolor}
\usepackage[footnotesize,labelsep=none,sl]{caption}
\usepackage{txfonts}
\usepackage{orcidlink}
\DeclareUnicodeCharacter{2212}{-}
\begin{document} 
\titlerunning{Gravity torques on Milky Way twins}
\authorrunning{P. da Silva and F. Combes}

   \title{Multiple-scale gas infall through gravity torques on Milky Way twins}

   \author{Patr\'icia da Silva
          \inst{1,2}\orcidlink{0000-0001-8837-8670}
          \and
          F. Combes\inst{1,3}\orcidlink{0000-0003-2658-7893}
          }

   \institute{ Observatoire de Paris, LERMA, PSL University, CNRS, Sorbonne University, Paris, France
         \and
             Instituto de Astronomia, Geof\'isica e Ci\^encias Atmosf\'ericas, Departamento de Astronomia, Universidade de S\~ao Paulo, S\~ao Paulo, Brazil\\
             \email{patricia2.silva@alumni.usp.br}
        \and
            Coll\`ege de France, 11 Pl. Marcelin Berthelot, 75231 Paris, France\\
              \email {francoise.combes@obspm.fr}
             }

   \date{Received May, 2024; accepted June, 2024}
 
  \abstract{
One of the main problems raised by the feeding of supermassive black holes (SMBHs) at the centres of galaxies is the huge angular momentum of the circumnuclear gas and of the gas reservoir in the galaxy disk. Because viscous torques are not efficient at kiloparsec or 100~pc scales, the angular momentum must be exchanged through gravity torques that arise from the non-axisymmetric patterns in the disks. Our goal here is to quantify the efficiency of bars and spirals in driving the gas towards the centre at different scales in galaxies. We selected a sample of nearby galaxies considered to be analogues of the Milky Way, that is, galaxies of late morphological type Sbc. Their bar strength was variable, either SB, or SAB, or SA, so that we were able to quantify the influence of the bar. The gravitational potential was computed from deprojected red images, either from Hubble Space Telescope or Legacy survey, depending on the spatial resolution and field of view considered. The torques were computed on the gas through CO emission maps from ALMA at different resolutions. H$\alpha$ maps from MUSE were used, when available. Eight out of ten galaxies are barred. The torques are found to be negative in the eight barred objects at kiloparsec scales, between corotation and the inner Lindblad resonance (ILR), with a loss of angular momentum in a few rotations. Inside the ILR, the torques are negative in only five cases, with a timescale of one to two rotations. The torques are positive for the galaxies without bars. The torques applied on the ionized gas are comparable to what is deduced from molecular gas. The bars are confirmed to be the essential pattern in the SMBH feeding at kiloparsec and 100~pc scales; higher-resolution gas maps are required to explore scales of 10~pc.
  }
   \keywords{Galaxies: evolution --  Galaxies: spiral  --  Galaxies: kinematics and dynamics  --
Galaxies: nuclei --  Galaxies: individual: ...}
   \maketitle

\section{Introduction}

\begin{figure*}[!ht]
\begin{center}

  \includegraphics[scale=0.21]{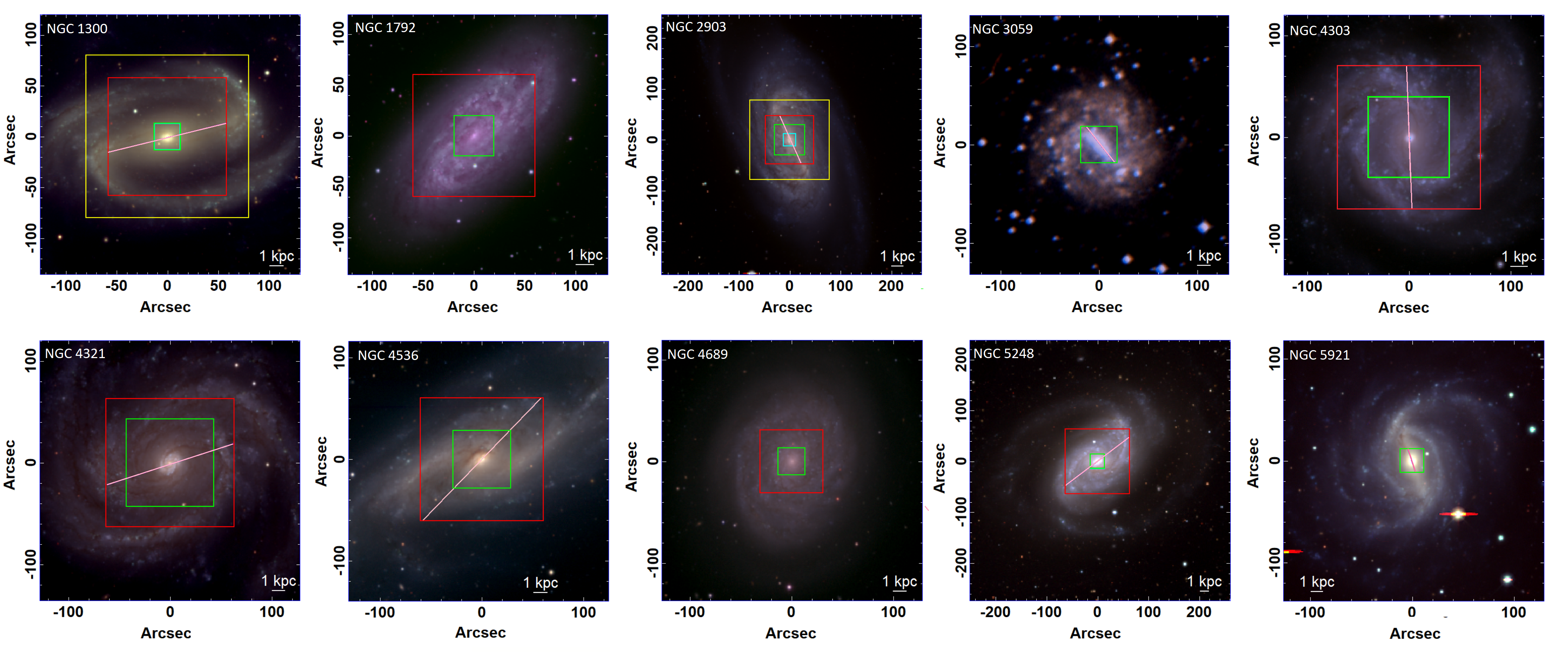}
  \caption{- Composite of Legacy images of the G (blue), R (green), and Z (red) bands (except for NGC 3059, which lacks Legacy images; in this case, we used the images from the Digitalized Sky Survey). The squares represent the size of the FOV that was used for each data set: green for HST and ALMA, cyan for the HST and the ALMA high spatial resolution set, red for Legacy and ALMA, and yellow for the Legacy and ALMA larger FOV (combination of two FOVs). The pink lines represent the PA of the bar. The orientation N-E is north up and east left.   \label{galaxiesimage}}
  
\end{center}
\end{figure*}

Supermassive black holes (SMBHs) are ubiquitous at the centres of galaxies, and they play an important role in their evolution. A possible correlation exists between the mass of the black hole and the bulge mass, suggesting a co-evolution of the two \citep{Kormendy2013}. One of the main phenomena invoked for this co-evolution is the feedback of active galaxy nucleus (AGN) episodes on the star formation rate in galaxies. AGN winds are able to produce molecular outflows from galactic nuclei, depleting the gaseous medium and moderating or quenching star formation in disks \citep{Fiore2017}. The efficiency of AGN winds is related to the Eddington ratio, and it is only significant when this ratio exceeds 0.01 \citep{Heckman2014}. Below this, the AGN feedback is more efficient in the radio mode through radio jets \citep{Garcia-Burillo2014, Dasyra2015}. 

Driving gas towards the galaxy centre to fuel both the central black hole and central star formation is an important ingredient in the co-evolution as well. The mechanisms for these inflows widely depend on the scale: at kiloparsec (kpc) scales in galaxy disks, viscous torques are negligible \citep{Lin1987}, and only gravity torques due to non-axisymmetric potentials are possible \citep{Buta1996}. Stellar bars are the most efficient and long-lived features in this domain, and they are able to remove the angular momentum of the gas in a few rotations \citep{Garcia-Burillo2005}. By computing gravity torques from the observed images of the old stellar component, together with the observed gas maps in nearby galaxies, \cite{Garcia-Burillo2012} concluded that the gas is driven inwards towards the SMBH in only 30\% of cases, while it is either stalled in a nuclear ring at the inner Lindblad resonance ({ILR}) of the bar or driven outwards, from corotation to the outer Lindblad resonance ({OLR}). The gas infall requires multi-scale mechanisms to reach the very centre, as was already described by \cite{Shlosman1989,Shlosman90,Berentzen2007}. For example, recent high-resolution ALMA observations have revealed that embedded nuclear bars efficiently feed the nucleus at a scale of 10 pc \citep{combes14, audibert19}. Simulations of nuclear bars that feed the nucleus were made by \citet{Englmaier2004}, and \citet{Laine2002} observed that more Seyfert galaxies are present in barred galaxies, proving the importance of this feature in the feeding of the nucleus.

We focus on spiral galaxies of the same type as the Milky Way (MW), that is, with an Sbc morphology, with different bar strengths (SABbc and SBbc) and non-barred counterparts. The study of morphological twins of the MW will allow us to understand the history of its secular evolution. Although our MW is the best-known galaxy because Earth lies in it, it is necessary to study its nearby twins for a grasp on the evolution it may have had and to understand its formation through gas infall, fostered by the bar. Time is very important because the environment can change drastically. For instance, the bar strength is likely not to be a long-lived characteristic (1 to 2 Gyr if there is gas in the disk) because it varies according to gas flows \citep{Bournaud2002} or even according to many other mechanisms during cosmological evolution \citep{Bi2022}. Regarding the AGN activity, we know that the Eddington ratio of the MW nucleus is very low $\sim$ 10$^{-9}$, but it could have been quite high in the past \citep{Sarkar2023}. When we take into account that AGN episodes are short-lived in galaxies, with a duty cycle of about 40 Myr \citep{Shankar2009}, the influence must be acknowledged as variable, if it exists. 

We study the molecular gas torques in order to understand the feeding processes in the MW analogues, and also compared with the non-barred galaxies to quantify the effect of the bar strength. By using data from Atacama Large Millimeter/Submillimeter Array (ALMA), Hubble Space Telescope (HST) images in the F814W filter, and the Legacy survey images in the Z and R bands, we were able to map the torque of the gas and to understand better how the molecular gas is driven towards the nuclear region of these galaxies. 

Section \ref{sample} presents the sample and the selection criteria, and  
Section \ref{methodology} describes the method we used to compute gravity torques in the galaxies. We analyse the data in Sec.~\ref{data}, and the results are discussed in Sec. \ref{disc}. Sec. \ref{conclusion} gathers out conclusions.

\begin{table*} [!ht]
\caption{- Observational information of the sample galaxies.}\label{obsinfo}
\resizebox{\textwidth}{!}{\begin{tabular}{ccccccccc}
\hline
         & \multicolumn{3}{c}{HST}                                                                                                                            & Legacy DR9     & \multicolumn{3}{c}{ALMA}                                                                                                                                                                                                                                                                  \\ \hline
Galaxy   & Proposal ID          & \begin{tabular}[c]{@{}c@{}}Observation \\ Date\end{tabular} & \begin{tabular}[c]{@{}c@{}}Exposition\\ Time (s)\end{tabular} & Proposal ID & Proposal ID                                                                                                                                    & \begin{tabular}[c]{@{}c@{}}Beam Major Axis\\ (arcsec)\end{tabular} & \begin{tabular}[c]{@{}c@{}}Rest\\ Frequency (MHz)\end{tabular}      \\ \hline
NGC 1300 & 15133 (PI: P. Erwin) & 10/27/2017                                                  & 500                                                           & 2012B-0001  & \begin{tabular}[c]{@{}c@{}}2015.1.00925.S (PI: G. Blanc)\\ 2018.1.01651.S (PI: A. Leroy)\\ 2019.1.01718.S (PI: S. Longmore)\end{tabular}       & \begin{tabular}[c]{@{}c@{}}7.45, 8.6\\ 1.99\\ 0.65\end{tabular}    & \begin{tabular}[c]{@{}c@{}}230.538\\ 230.538\\ 345.796\end{tabular} \\
NGC 1792 & 15654 (PI: J. Lee)   & 05/25/2019                                                  & 830                                                           & 2012B-0001  & 2017.1.00886.L (PI: E. Schinnerer)                                                                                                             & 1.92                                                               & 230.538                                                             \\
NGC 2903 & 9788 (PI: Ho, L.)                & 03/01/2004                                                  & 120                                                           & 2014B-0404  & \begin{tabular}[c]{@{}c@{}}2017.1.00886.L (PI: E. Schinnerer)\\ 2018.1.00517.S (PI: K. Onishi)\\ 2019.1.01730.S (PI: J. Puschnig)\end{tabular} & \begin{tabular}[c]{@{}c@{}}1.99\\ 0.63\\ 4.64, 3.69\end{tabular}   & \begin{tabular}[c]{@{}c@{}}230.538\\ 230.538\\ 345.800\end{tabular} \\
NGC 3059 & 9042 (PI: Smartt, S.)                 & 08/13/2001                                                  & 160                                                           & --          & 2017.1.00886.L (PI: E. Schinnerer)                                                                                                             & 1.99                                                               & 230.538                                                             \\
NGC 4303 & 15654 (PI: J. Lee)   & 03/29/2020                                                  & 803                                                           & 2014B-0404         & 2015.1.00956.S (PI: A. Leroy)                                                                                                                  & 1.99                                                               & 230.538                                                             \\
NGC 4321 & 15654 (PI: J. Lee)   & 03/15/2020                                                  & 836                                                           & 2014B-0404  & 2015.1.00956.S (PI: A. Leroy)                                                                                                                  & 1.99                                                               & 230.538                                                             \\
NGC 4536 & 15654 (PI: J. Lee)   & 03/22/2020                                                  & 803                                                           & 2014B-0404          & 2017.1.00886.L (PI: E. Schinnerer)                                                                                                             & 1.99                                                               & 230.538                                                             \\
NGC 4689 & 15133 (PI: P. Erwin) & 12/22/2017                                                  & 500                                                           & 2014B-0404  & 2017.1.00886.L (PI: E. Schinnerer)                                                                                                             & 1.99                                                               & 230.538                                                             \\
NGC 5248 & 15133 (PI: P. Erwin) & 01/04/2018                                                  & 500                                                           & 2014B-0404  & \begin{tabular}[c]{@{}c@{}}2017.1.00886.L (PI: E. Schinnerer)\\ 2019.1.00876.S (PI: E. Schinnerer)\end{tabular}                                & \begin{tabular}[c]{@{}c@{}}1.99\\ 0.29\end{tabular}                & \begin{tabular}[c]{@{}c@{}}230.538\\ 230.538\end{tabular}           \\
NGC 5921 & 15323 (PI: J. Walsh) & 09/13/2017                                                  & 738                                                           & --          & 2019.1.01742.S (PI: D. Rosario)                                                                                                                & 0.49                                                               & 230.538                                                             \\ \hline
\end{tabular}}
\end{table*}

\begin{table*}[!ht]
\centering
\caption{- Properties of the sample galaxies.}\label{tabelagal}
\resizebox{\textwidth}{!}{\begin{tabular}{ccccccccc}
\hline
Galaxy   & \begin{tabular}[c]{@{}c@{}}Morphological \\ Type\end{tabular} & Distance (Mpc) & Nuclear Activity     & PA ($^{\circ}$) & Bar PA ($^{\circ}$) & Inclination ($^{\circ}$) & \begin{tabular}[c]{@{}c@{}}Total apparent corrected\\ B-magnitude (mag)\end{tabular}  & \begin{tabular}[c]{@{}c@{}} Total Stellar Mass\\ log($M_{*}/M_{\bigodot}$)\end{tabular} \\ \hline
NGC 1300 & SB(rs)bc  & 14.5  & LINER (AGN)  & 278$^{a}$   &104$^{{1}}$  & 31.8$^{a}$ & 10.61 & 10.30$^{{7}}$  \\
NGC 1792 & SA(rs)bc   & 11.3   & HII Region         & 316  &    --     & 63.1      & 10.18 & 10.57$^{{8}}$   \\
NGC 2903 & SAB(rs)bc  & 8.57  & HII Region & 202   &   24$^{{2}}$      & 67.1    & 8.83 & 10.31$^{{9}}$  \\
NGC 3059 & SB(rs)bc  & 14.8   &  HII Region   & 330   &   39$^{{3}}$    & 21.3   & 10.63 &  10.40$^{{8}}$  \\
NGC 4303 & SAB(rs)bc & 12.2  & Seyfert 2/LINER  & 310   &   2$^{{4}}$   & 18.1   & 10.02 & 10.48$^{{7}}$ \\
NGC 4321 & SAB(s)bc & 15.8& LINER (AGN) & 156$^{a}$    &   108$^{{4}}$   & 24    & 9.83 & 10.71$^{{10}}$ \\
NGC 4536 & SAB(rs)bc  & 14.7 & LINER (AGN) & 306$^{a}$     & 136$^{{5}}$   & 66$^{a}$   & 10.27 & 10.19$^{{10}}$ \\
NGC 4689 & SA(rs)bc  & 15.7  & HII Region   & 163    &   ---     & 42.5     & 11.23 & 10.58$^{{8}}$  \\
NGC 5248 & SAB(rs)bc  & 11.8  & Seyfert/LINER  & 120  &  127$^{{4}}$  & 56.4     & 10.39  & 10.30$^{{11}}$  \\
NGC 5921 & SB(r)bc  & 17.3   &  HII Region    & 137  &    18$^{{6}}$     & 49.5    & 11.2 & 10.41$^{{12}}$ \\ \hline 
\end{tabular}}
{\fontsize{7}{9}{\textit{Notes.} The distances are medians of the values listed on NED and morphological types are from \citet{RC3}. PA, Inclination, and the Total apparent corrected B band magnitude were obtained from Hyperleda \citep{hyperleda}. The values for the PA and inclination of NGC 1300, NGC 4321 and NGC 4536 (indicated by "$a$") are from \citet{Lang20}. The nuclear activity classifications are discussed in Appendix \ref{revisiongalaxies}.{References for the bar PA: (1) \citet{Lindblad97}, (2) \citet{Sheth02}, (3) \citet{bottema90}. (4) \citet{das03}, (5) \citet{shaw95}, (6) \citet{martin97}. References for the total stellar mass: (7) \citet{davis19}, (8) \citet{weinzirl09}, (9) \citet{murugeshan20}, (10) \citet{zabel22}, (11) \citet{Chiang24}, (12) \citet{diaz-garcia21}.}}}
\end{table*}

\section{Sample} \label{sample}

\begin{figure*}[!ht]
\begin{center}

  \includegraphics[scale=0.21]{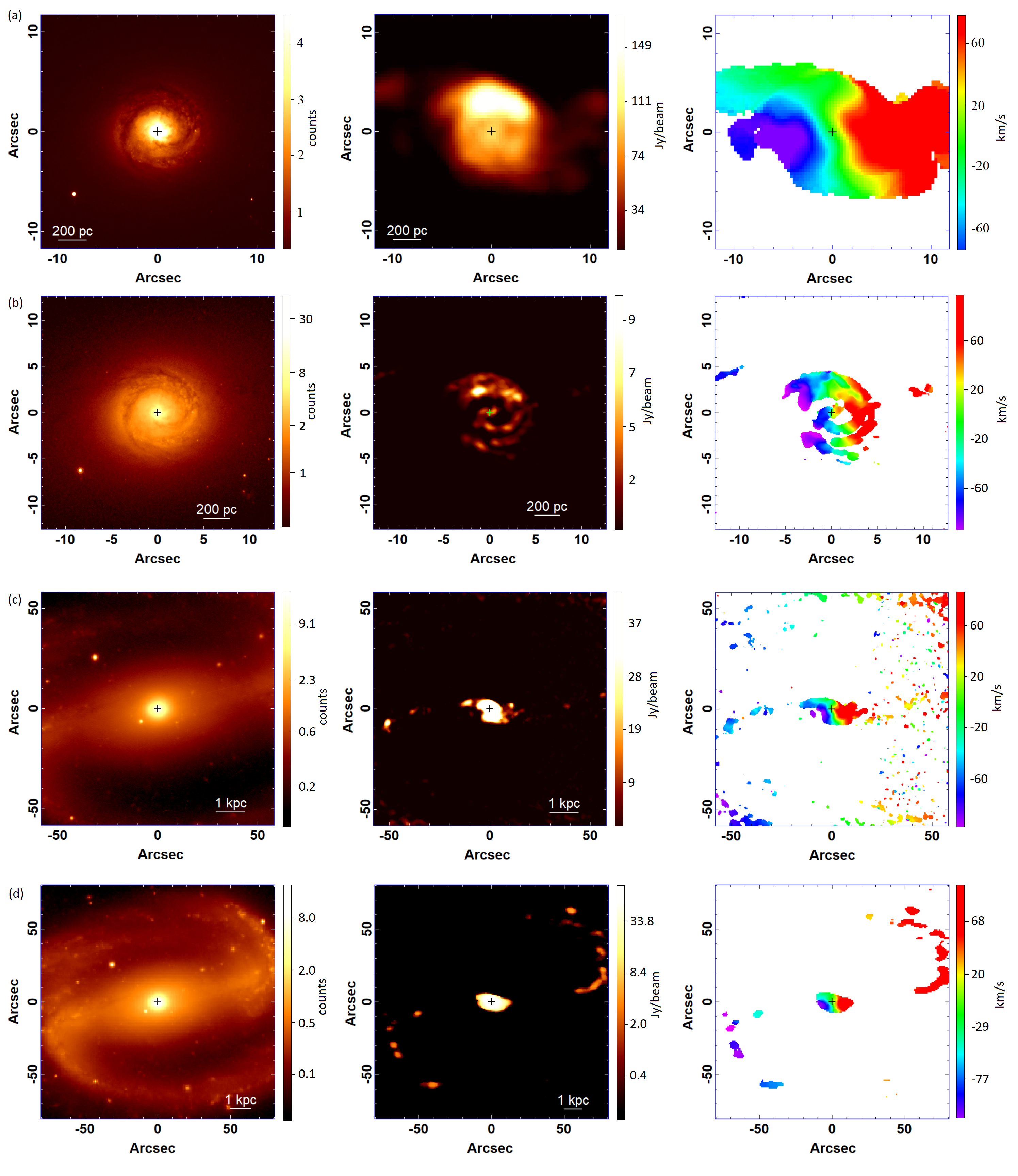}
  \caption{- NGC 1300. Left panel: Images of the F814W filter of HST (a, b),  and Legacy Z band (c, d). Middle and right panels: ALMA data (mean flux and velocity map) associated with each FOV, as described in section \ref{methodology}. The orientation N-E is north up and east right. The cross (arbitrary size) indicates the centre of the image, which was taken from the emission peak from the infrared image and was used to match the ALMA images. \label{NGC1300_m0_m1}}
  
\end{center}
\end{figure*}

 In order to estimate the gravitational potential in the disk, we needed red images of the galaxies to avoid dust extinction and to avoid biasing by the light of a small number of OB stars with low mass-to-light ratios (M/L). The red images were available with several resolutions and fields of view (FOVs): We used HST images in the F814W filter and Legacy survey DR9 images in the R (for NGC 4303 and NGC 4536 because their Z-band images are saturated at the nucleus) and Z bands. To compute the torques on the molecular gas, we used ALMA data mainly from the Phangs survey \citep{PHANGS} and from the ALMA archive. The data details are displayed in Table \ref{obsinfo}.

We searched for the data of SABbc, SBbc, and SAbc galaxies with inclinations lower than 65$^{\circ}$ because the computations cannot be performed for edge-on galaxies. We also only selected nearby galaxies (D$<$ 18 Mpc) to have enough spatial resolution, and galaxies with a magnitude in the B-band lower than 12 for a complete sample. Based on the classification in the NASA/IPAC Extragalactic Database (NED\footnote{The NASA/IPAC Extragalactic Database (NED) is funded by the National Aeronautics and Space Administration and operated by the California Institute of Technology.}), these criteria led to 43 galaxies. The availability of the data limited the sample to ten galaxies, with the exceptions of the two galaxies that were already studied (NGC 1566 \citealt{combes14} and NGC 613 \citealt{audibert19}). Another two galaxies have ALMA data, but the FOV was too small for this work : NGC 1255 and NGC 4030, and there were problems with the data of a final two galaxies: NGC 3521 and NGC 5530. When these galaxies were removed, we obtained ten galaxies. Their properties are shown in Table \ref{tabelagal}. The inclinations of NGC 2903 and NGC 4536 are higher than 65$^{\circ}$. However, given the uncertainties on this parameter, which is about 5$^{\circ}$ at least, we chose to include these galaxies in the sample.

\begin{figure*}[!ht]
\begin{center}

  \includegraphics[scale=0.21]{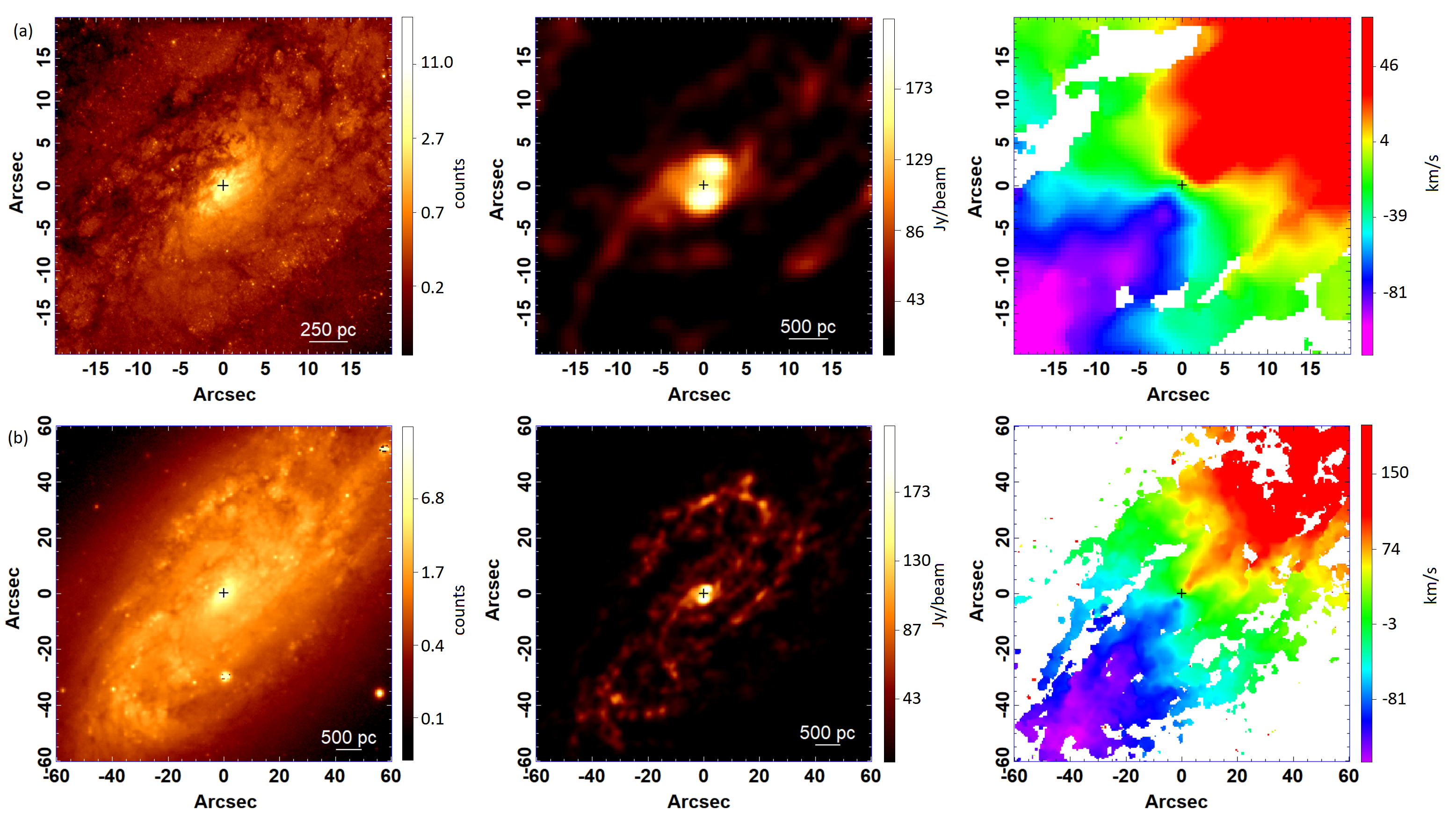}
  \caption{- NGC 1792. Same as the left panel of Fig.~\ref{NGC1300_m0_m1}. (a) Image of the HST filter F814W and (b) image of the Z band from the Legacy survey. \label{NGC1792_m0_m1}}
  
\end{center}
\end{figure*}

\begin{figure*}[!ht]
\begin{center}

  \includegraphics[scale=0.21]{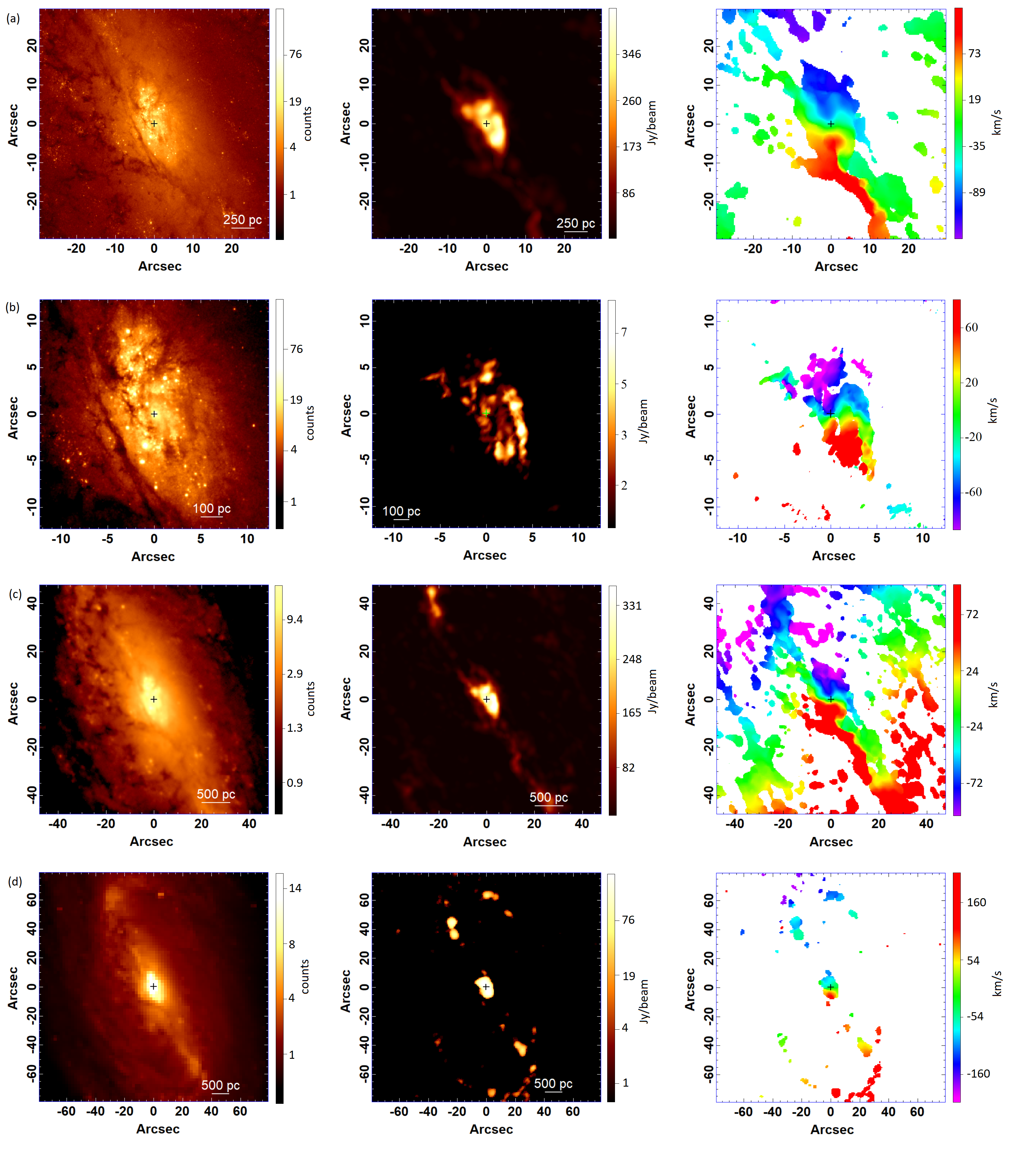}
  \caption{- NGC 2903. Same as Fig.~\ref{NGC1300_m0_m1} \label{NGC2903_m0_m1}}
  
\end{center}
\end{figure*}

\section{Method}\label{methodology}

The goal of this work is to quantify the rate of gas inflow
due to gravity torques exerted by a bar (or spirals) on the molecular gas. This work was done at different scales, according to the spatial resolutions and FOVs of the observations.

The gravitational potential is mainly due to stars, and because the old population is more representative of the mass, red or near-infrared images are the best tracers, while the dark matter can be neglected in the central parts of MW-like galaxies 
\citep{Persic1996}. The red images at high resolution are usually from the filter F814W from HST for the central region and from the Legacy survey (at bands Z or R) at large scales to benefit from a wider FOV, but the spatial resolution is lower. 

The images were rotated and deprojected according to the PAs and inclination angles given in Table \ref{tabelagal}. The bulge component was separated for the deprojection, because it was assumed to be spherical; in general, the bulge does not significantly contribute to the mass, because of the late morphological types of the targets. After rotation and deprojection, the images were Fourier transformed to compute the gravitational potential and forces. The M/L was assumed to be constant throughout the disk and bulge. Its precise value was considered not to be important because we computed normalised torques, corresponding to the relative loss of angular momentum per rotation. To take the third dimension of the disk into account, we assumed a stellar exponential disk thickness of $\sim$1/12th of the radial scale length of the galaxies
\citep{Garcia-Burillo2005, audibert19}. After computing the potential and forces in each pixel, we computed the torques with an emission map of the molecular gas at the same resolution. All maps were resampled to the same number of pixels and pixel sizes. 

The amount of data available for both infrared images and molecular gas varies for a given galaxy. NGC 1300 has a considerable volume of data. In this case, we opted to use four ALMA data sets: two data sets were from the program 2015.1.00925.S and were combined in order to form a larger FOV (represented by the yellow square in Fig.~\ref{galaxiesimage} of NGC 1300). These data were used in the analysis with a Legacy image with a corresponding FOV. We used Phangs ALMA data associated with HST and Legacy images (green and red squares) and also the ALMA data from the program 2019.1.01718.S, which has a better spatial resolution, associated with the same HST image (cyan square). For NGC 1792, we used Phangs data together with HST (green square in Fig.~\ref{galaxiesimage} of NGC 1792) and Legacy images (red square). NGC 2903 also had four data sets: two data sets from the program 2019.1.01730.S, which had the same goal as NGC 1300. They were combined to form a larger FOV and were used together with a Legacy image that was matched to have the same size (yellow square). This galaxy also had data from Phangs that were used together with Legacy and HST data (red and green squares, respectively). We also used high spatial resolution ALMA data (program 2018.1.00517.S) together with HST data represented by the cyan square in Fig.~\ref{galaxiesimage} of NGC 2903. For NGC 3059, NGC 4303, NGC 4321, NGC 4536, and NGC 4689, only Phangs ALMA data were used, together with HST and Legacy data (with the exception of NGC 3059, which lacks Legacy images). For NGC 5248, the Phangs data were used together with HST and Legacy images as well; however, ALMA data were also used from the program 2019.1.00876.S (with higher spatial resolution) together with the HST images. Finally, for NGC 5921, we used ALMA data from 2019.1.01742.S program with a high spatial resolution, together with the HST image of this galaxy. This galaxy has Legacy images; because the ALMA data have a small FOV, it would provide the same results as were obtained with HST images with a lower spatial resolution, however. For better results, we used HST data. The summary of all information is shown in Table~\ref{obsinfo} and Fig.~\ref{galaxiesimage}.

To prepare the data for the computation of gravitational torques from the infrared images, we first removed the MW bright stars in the FOV.
We then rotated the images using the PA of the galaxy (Table~\ref{tabelagal}) so that its major axis was horizontal. This implied the definition of a common centre.
Using the infrared image, we established a centre based on the emission peak of the nuclear source. We also centred the ALMA data cube according either to the emission peak or to the centre of the ring, and we used the velocity field. The two components (infrared images and ALMA data cube) were then re-centred so that they had the same centre. The images were then deprojected by stretching the vertical axis.
  For the ALMA data, the same rotation and deprojection were applied to the data cube converted into velocities relative to the centre. These processes were performed with scripts in Python and Gildas environments.

\section{Data analysis} \label{data}

\subsection{Gas morphology and kinematics}
\label{morphokin}

Fig.~\ref{NGC1300_m0_m1}(a) and (b) show the nuclear structure of NGC 1300. The morphology is that of a ring with a radius of about 300~pc (already detected in previous works; see Appendix \ref{revisiongalaxies}) , which is decomposed into two concentric rings in the ALMA mean image. There is clear evidence of a structure that connects the nuclear source to the ring.
The nuclear low-resolution image reveals that this ring is connected to the spiral structure. This is better visible in the images with a larger FOV (Fig.~\ref{NGC1300_m0_m1}c and d). The ring and spiral obey the same regular rotation. The two spiral arms emerge outside the ring. They first correspond to the leading dust lanes in the bar, and then continue to wind out outside the bar.

Fig.~\ref{NGC1792_m0_m1}(a) and (b) show NGC 1792 data in two different FOVs. They show the nuclear region, which is considerably flocculent. The nuclear source appears to be divided into two regions (as seen in the ALMA data). Because of the spatial resolution, however, we cannot conclude this with certainty. This galaxy has a short nuclear bar that might be related to this structure (see Appendix \ref{revisiongalaxies}). The gas clearly rotates in both FOVs. The spiral arms from the galaxy appear to be connected to the nuclear source seen in the nuclear ALMA image. This galaxy is classified as non-barred, but these maps are quite similar to those of the barred counterparts.

The structure of the nuclear region of NGC 2903 (Fig.~\ref{NGC2903_m0_m1}) is very flocculent and hard to interpret. However, a nuclear ring can be distinguished in the gas map. According to the literature, this ring has radius of about 600 pc (see Appendix \ref{revisiongalaxies} for more information). This structure is also connected to the spiral arms of the galaxy and follows the orientation of the galaxy bar (clearly visible in panel d). The gas obeys a clear rotation, and, as in NGC 1300, the ring also follows the same rotation pattern as the remaining galaxy disk.

\begin{figure*}
\begin{center}

  \includegraphics[scale=0.21]{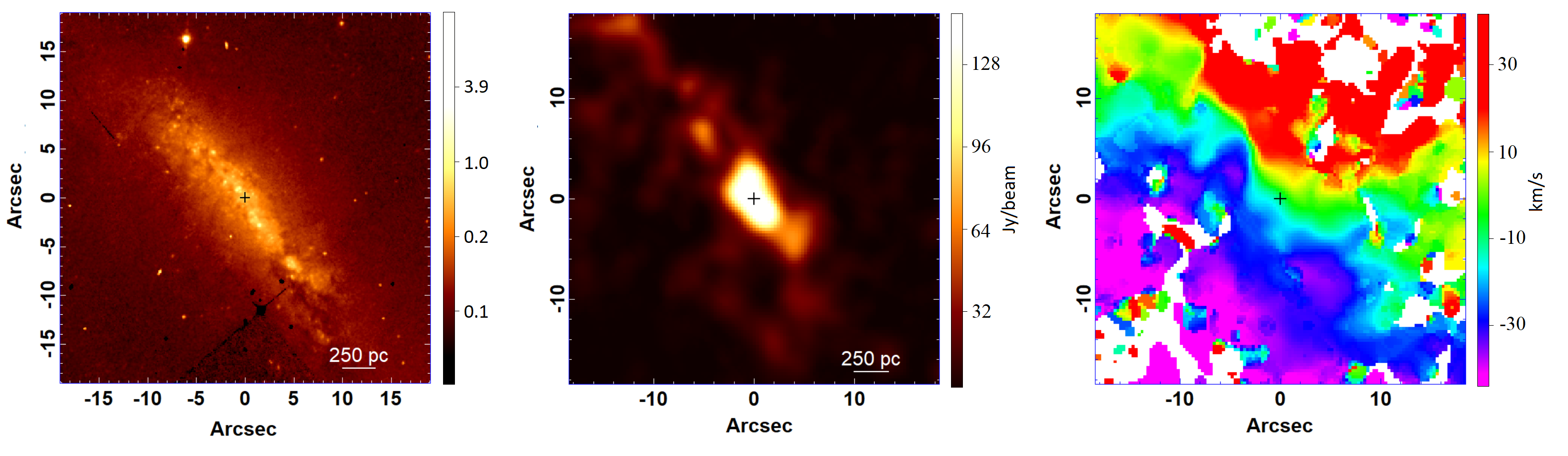}
  \caption{- NGC 3059. Same as Fig.~\ref{NGC1300_m0_m1}. The HST image is shown in the left panel. \label{NGC3059_m0_m1}}
  
\end{center}
\end{figure*}

The data from NGC 3059 (Fig.~\ref{NGC3059_m0_m1}) do not present much detail because of the spatial resolution; the bar is very thin and strong in this almost face-on galaxy. The nuclear source is connected to the spiral arms in the same direction as the bar. The rotation is also clear in this case. This galaxy was reported to have a weak bar (see Appendix \ref{revisiongalaxies}), but we found that this galaxy has the strongest bar in the sample. 

\begin{figure*}
\begin{center}

  \includegraphics[scale=0.21]{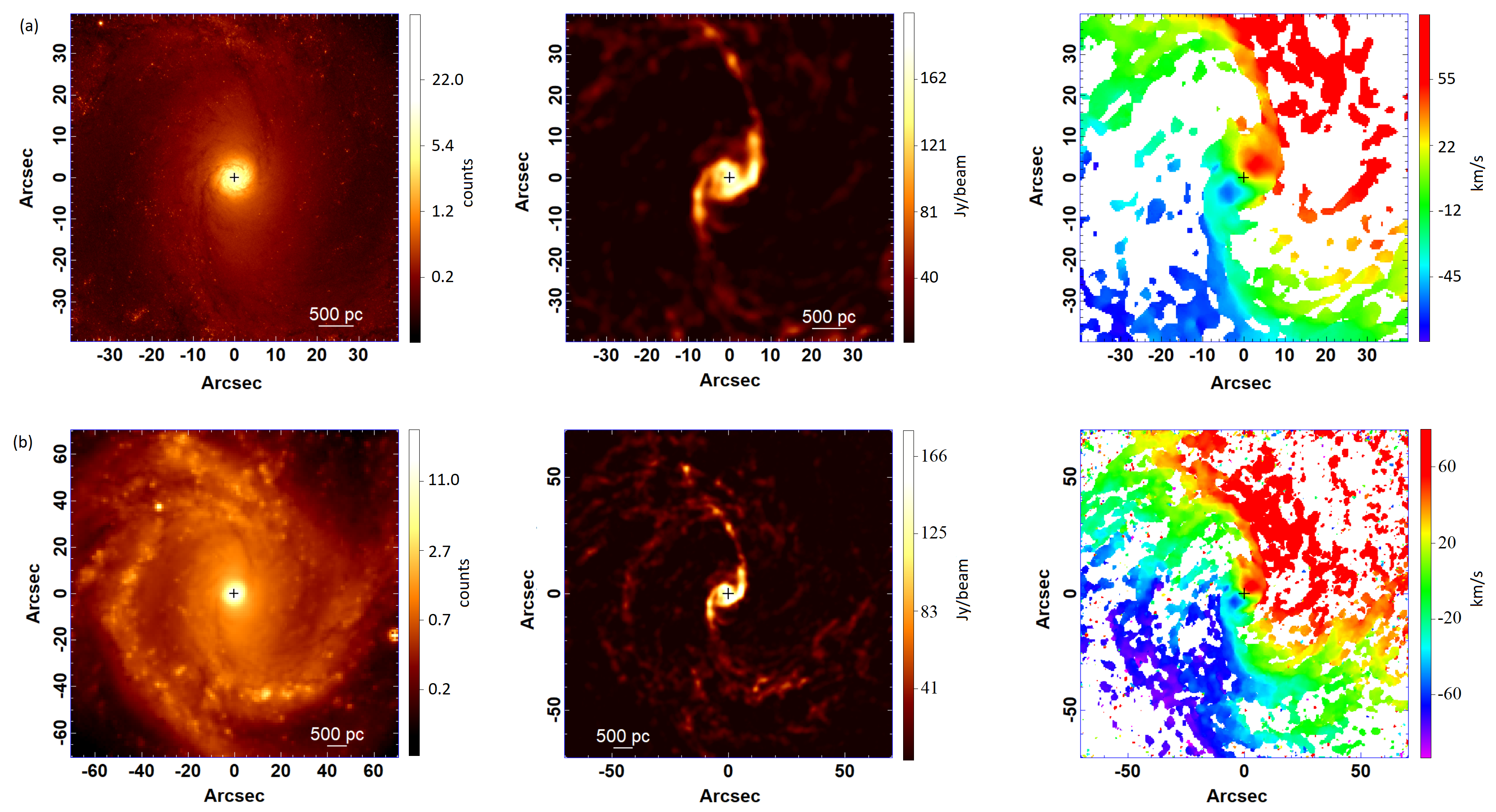}
  \caption{- NGC 4303. Same as Fig.~\ref{NGC1792_m0_m1}. In this case, an R-band image from the Legacy survey was used. \label{NGC4303_m0_m1}}
  
\end{center}
\end{figure*}

NGC 4303 (Fig.~\ref{NGC4303_m0_m1}) presents a nuclear ring at the bar ILR in the centre that is connected to the spiral arms, leading dust lanes along the bar. These in turn are connected to the spiral arms outside the bar. The ring radius is about 310 pc. The spiral arms are very clear and rotate regularly. The spatial resolution did not allow us to see the nuclear bar that was reported in the literature (see Appendix \ref{revisiongalaxies} for more details).

\begin{figure*}
\begin{center}

  \includegraphics[scale=0.21]{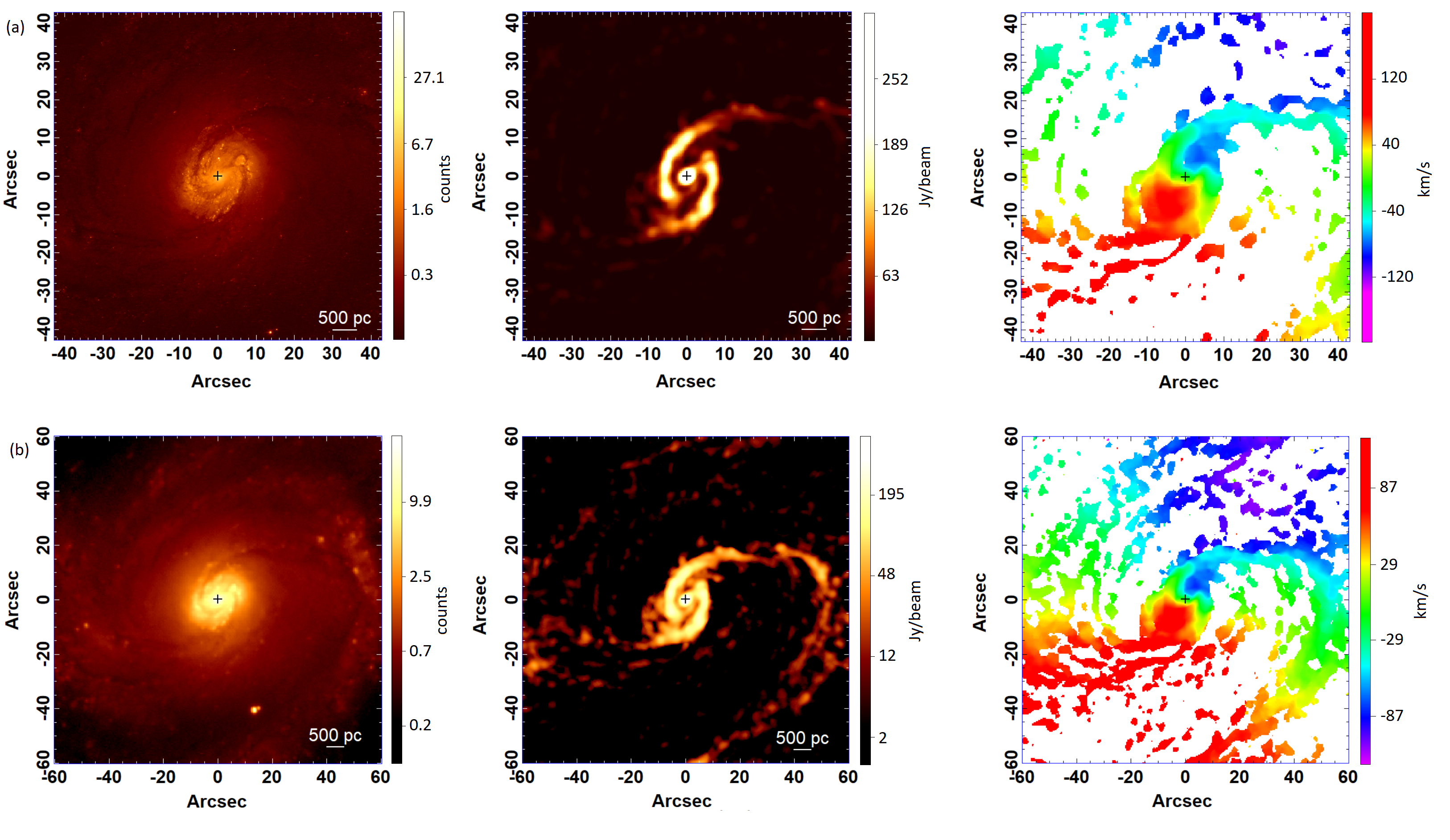}
  \caption{- NGC 4321. Same as Fig.~\ref{NGC1792_m0_m1}. \label{NGC4321_m0_m1}}
  
\end{center}
\end{figure*}

The gas morphology of NGC 4321 (Fig.~\ref{NGC4321_m0_m1}) is very similar to that of NGC 4303. The spiral arms follow the morphology typical of a barred galaxy. The nuclear ring radius is $\sim$ 500 pc. As in 4303, the spatial resolution was not high enough for us to detect the nuclear bar that was observed in previous studies (see Appendix  \ref{revisiongalaxies}).

\begin{figure*}
\begin{center}

  \includegraphics[scale=0.21]{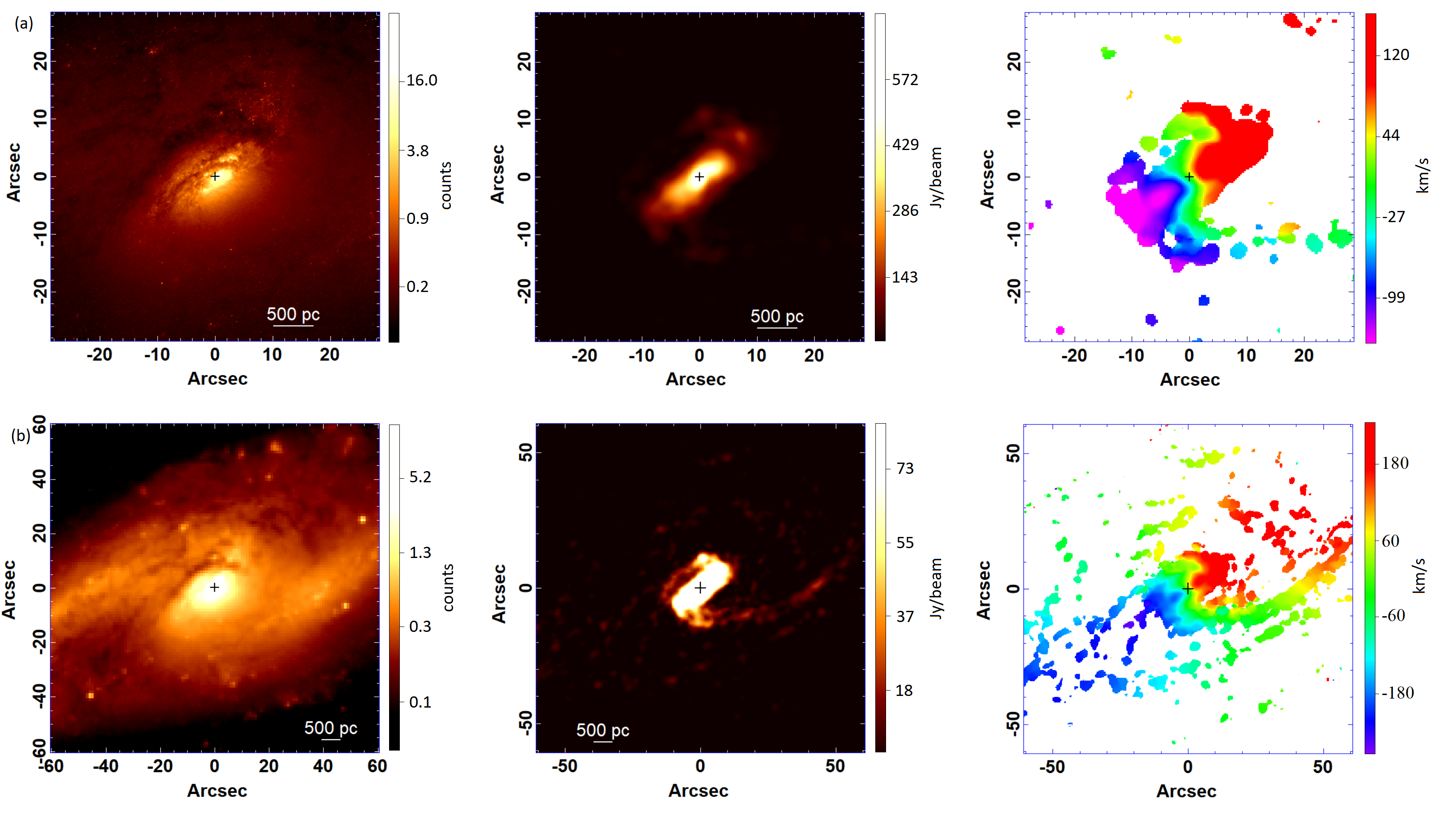}
  \caption{- NGC 4536. Same as Fig.~\ref{NGC1792_m0_m1}. In this case, an R-band image from the Legacy survey was used.\label{NGC4536_m0_m1}}
  
\end{center}
\end{figure*}

NGC 4536 has a very obscured nucleus, with a conspicuous dust lane on the near side, in the north-east. The bar orientation of this galaxy is confused with its major axis because of the high inclination. However, in Fig.~\ref{NGC4536_m0_m1}, molecular gas images show a clear nuclear bar with a PA~$\sim$~136$^{\circ}$ and a diameter of about 2.1 kpc that was also detected in previous studies (see Appendix \ref{revisiongalaxies}). One gas spiral arm is also connected to the bar in the south-west. The opposite arm in the north-east is better visible in the stellar component. 

\begin{figure*}
\begin{center}

  \includegraphics[scale=0.21]{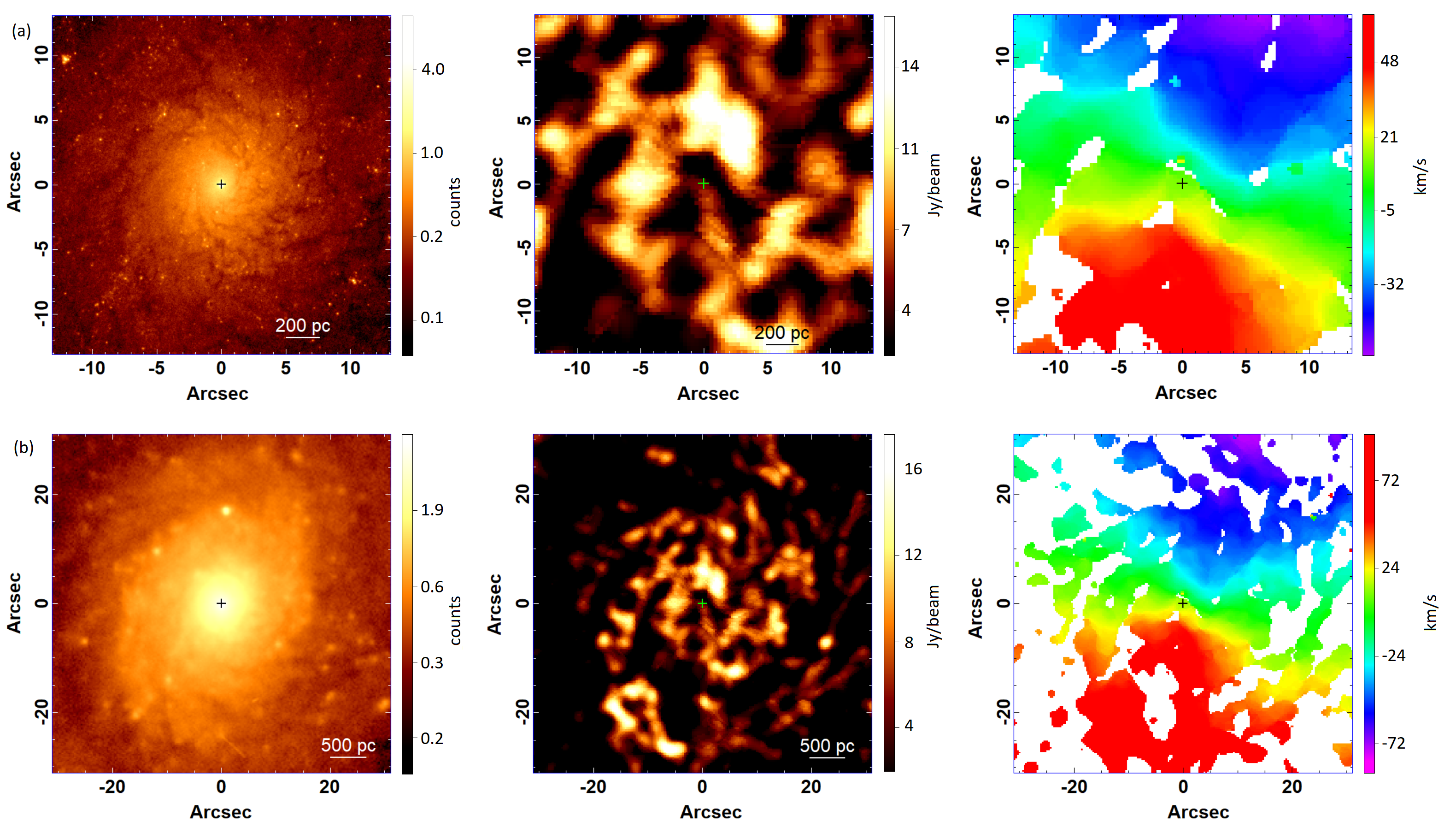}
  \caption{- NGC 4689. Same as Fig.~\ref{NGC1792_m0_m1}.\label{NGC4689_m0_m1}}
  
\end{center}
\end{figure*}

NGC 4689 is another non-barred galaxy. In this case, the difference between this non-barred galaxy and the barred galaxies is clear: The nucleus, which is very flocculent, lacks any clear structure (Fig.~\ref{NGC4689_m0_m1}). Some studies reported evidence of a weak bar, but we detected no clear structure in this galaxy (see Appendix \ref{revisiongalaxies}). The radio data show no clear nuclear source either. Moreover, the spiral structure in the molecular gas images is not as clear as in the infrared images. 

\begin{figure*}
\begin{center}

  \includegraphics[scale=0.21]{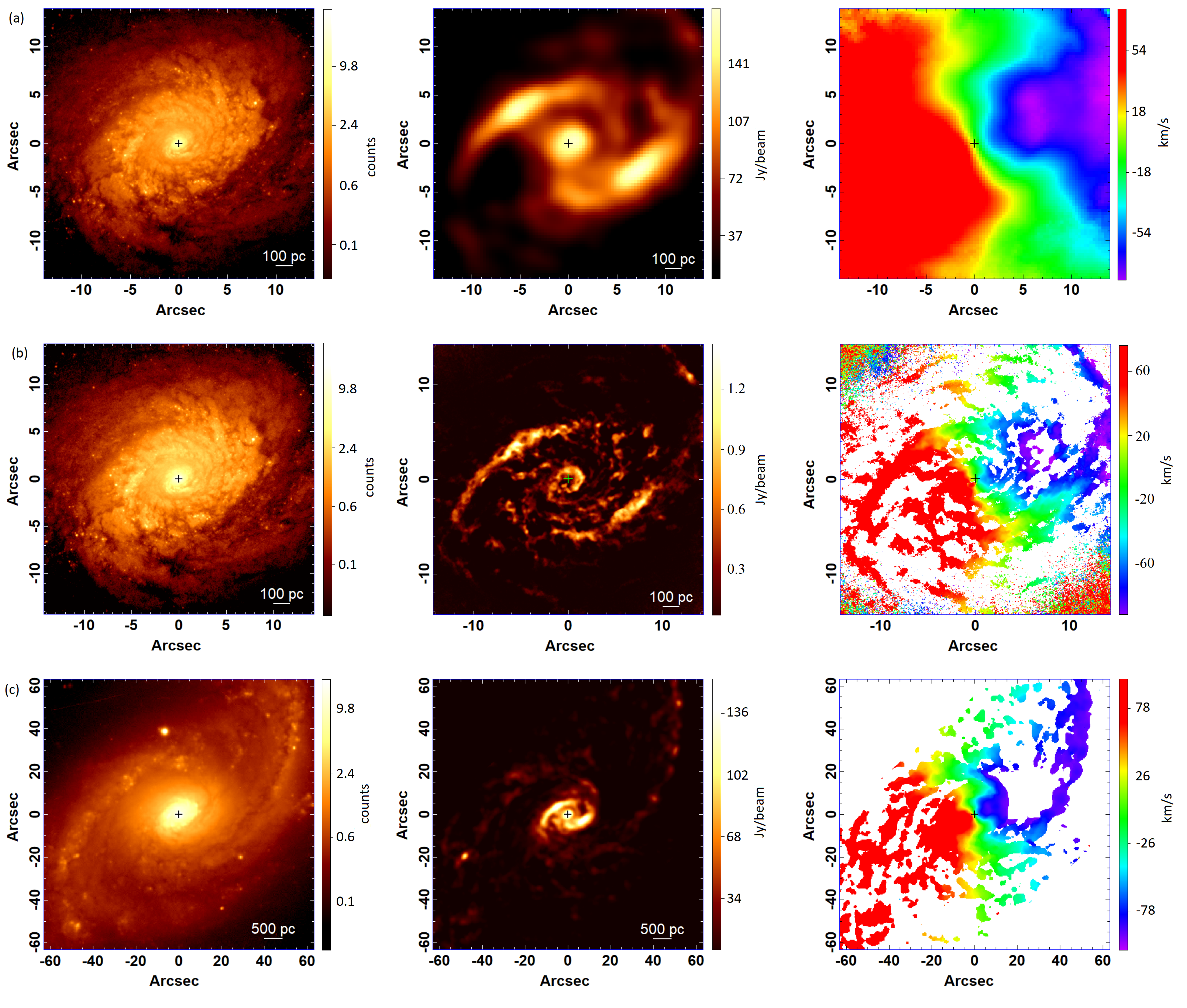}
  \caption{- NGC 5248. Same as Fig.~\ref{NGC1300_m0_m1}. Panels (a) and (b) show images of the HST filter F814W, and panel (c) shows an image from the Z band from the Legacy survey.\label{NGC5248_m0_m1}}
  
\end{center}
\end{figure*}

NGC 5248 has a very clear spiral structure and a nuclear ring at the centre, with a radius of about 88 pc (Fig.~\ref{NGC5248_m0_m1}). It also has an inner pseudo-ring that encircles the bar. This is expected to delineate the ultra-harmonic resonance near the corotation at a radius $\sim$ 500 pc. Then the spiral arms connect to this pseudo-ring outside the bar. We did not detect the possible nuclear bar (see Appendix \ref{revisiongalaxies}). This might be due to the low spatial resolution. 

\begin{figure*}
\begin{center}

  \includegraphics[scale=0.21]{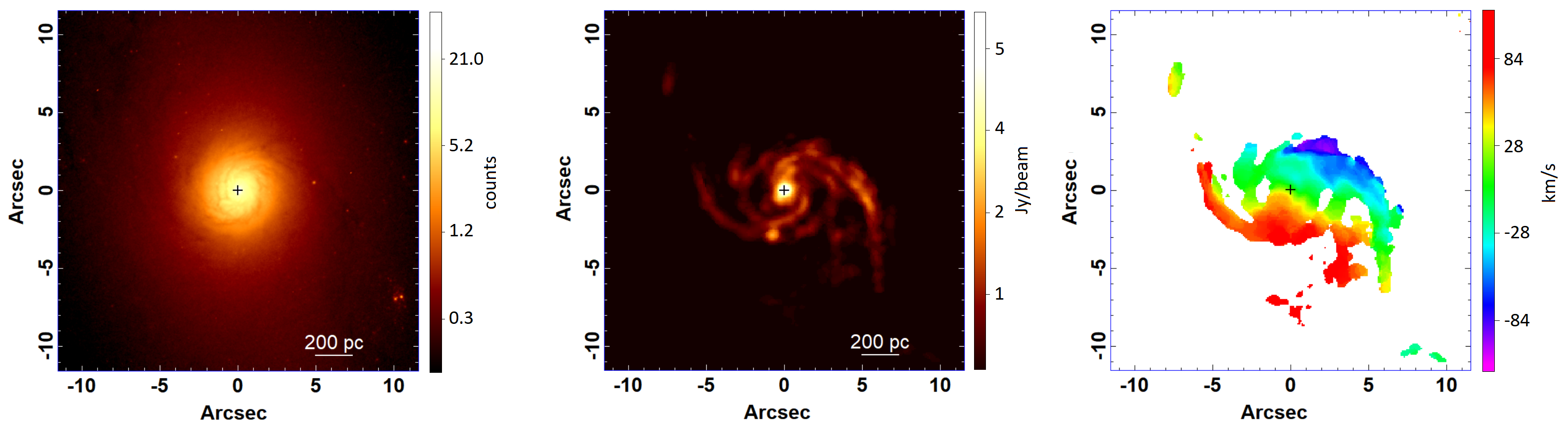}
  \caption{- NGC 5921. Left panel: F814W image from HST. Middle and right panels: ALMA data as in Fig.~\ref{NGC1300_m0_m1}. \label{NGC5921_m0_m1}}
  
\end{center}
\end{figure*}

Several structures are embedded in galaxy NGC 5921 (Fig.~\ref{NGC5921_m0_m1}). The morphology at large scales 
suggests that an inner ring encircles the bar with a radius $\sim$ 5 kpc. Two spiral arms are connected to this ring/bar (see Fig. \ref{galaxiesimage}). At smaller scales, a nuclear ring with a radius of about 230 pc is visible. Nuclear spiral arms are also connected to this nuclear source.

In summary, all barred galaxies of the sample present a very complex nuclear region. Six galaxies present clear evidence of nuclear rings (N1300, N2903, N4303, N4321, N5248, and N5921). Two objects have a nuclear bar (N4536 and N5921). NGC 4689 alone did not show clear structures in its nucleus. This might be related to the lack of a bar. Although it was classified as a non-barred galaxy, NGC 1792 has a clear structure in its nucleus.

\begin{figure}
\begin{center}

  \includegraphics[scale=0.5]{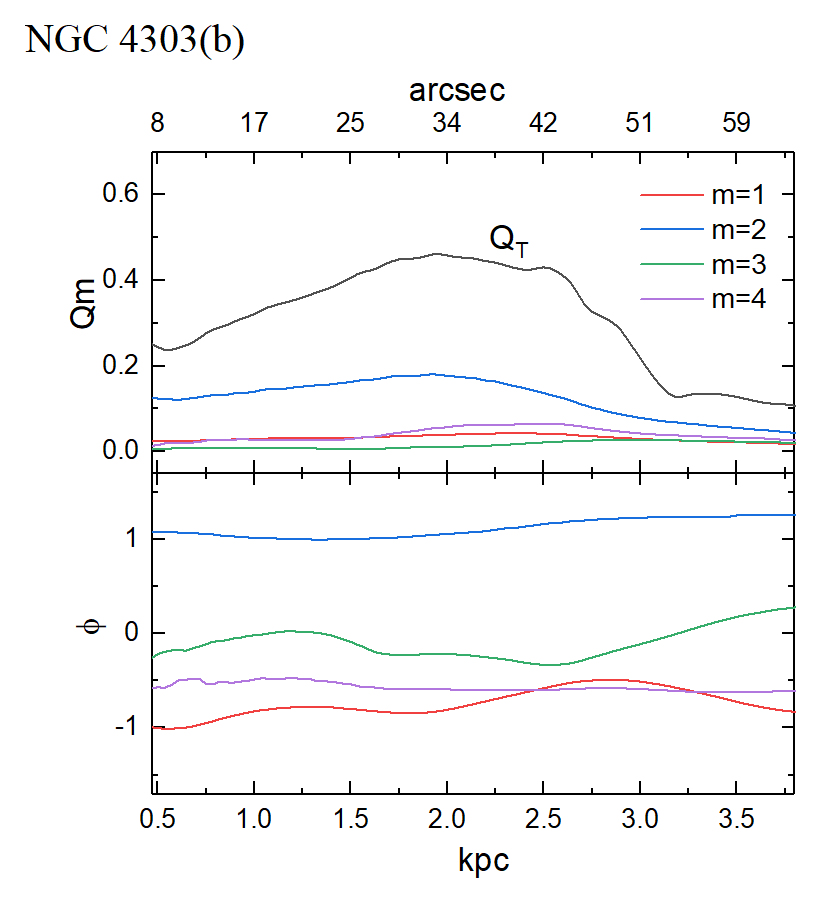}
  \caption{- Radial distribution of the amplitudes (top) and phases (bottom) of the non-axisymmetric perturbations in each Fourier component (m=1 to 4). Q$_T$ represents the total perturbation curve in the potential for the data set of N4303 (b). This plot illustrates the typical potential component curves obtained in the sample. The maximum values of the Q$_2$ and Q$_T$ curves are shown in Table~\ref{tab:qplot}. \label{fig:qplot}}
  
\end{center}
\end{figure}

\begin{table*}
\centering
\caption{-  Maximum and minimum values with their associated radii of the torque and relative maxima of non-axisymmetric potential curves ($Q_2$ and $Q_T$).}\label{tab:qplot}
\resizebox{0.7\textwidth}{!}{%
\begin{tabular}{ccccccc}
\hline
Galaxy        & Radius (kpc)                                                                    & \begin{tabular}[c]{@{}c@{}}dL/L\\ in one rotation\end{tabular}   & Radius (kpc)  & Qmax (m=2)                & Radius (kpc)      & Qmax  (Total)      \\ \hline
NGC 1300 (a)  & \begin{tabular}[c]{@{}c@{}}0.33\\ 0.53\end{tabular}                             & \begin{tabular}[c]{@{}c@{}} 0.025\\ -0.063\end{tabular}    & 0.5& 0.11 B & 0.234 & 0.061 R\\ \hline
NGC 1300 (b)  & \begin{tabular}[c]{@{}c@{}}0.083\\ 0.4\end{tabular}                            & \begin{tabular}[c]{@{}c@{}} 0.084\\ -0.237\end{tabular}   &  0.68   & 0.013 B   & \begin{tabular}[c]{@{}c@{}} 0.21\\ 0.747\end{tabular} & \begin{tabular}[c]{@{}c@{}} 0.034 R\\ 0.031 B \end{tabular}  \\ \hline
NGC 1300 (c)  & ---     & ---  &  2.66  &  0.143 B & 3.04   &  0.359 B  \\ \hline
NGC 1300 (d)  & ---     & ---    &  3.55  &  0.222 B & 3.55  &  0.567 B\\ \hline
NGC 1792 (a)  & \begin{tabular}[c]{@{}c@{}} 0.09 \\ 0.208\end{tabular}            & \begin{tabular}[c]{@{}c@{}}-0.145 \\ 0.148\end{tabular}   & \begin{tabular}[c]{@{}c@{}} 0.276\\ 0.473\end{tabular} & \begin{tabular}[c]{@{}c@{}} 0.265 S \\ 0.208  S\end{tabular} & \begin{tabular}[c]{@{}c@{}}0.276\\ 0.473 \\ 0.83\end{tabular}  & \begin{tabular}[c]{@{}c@{}}0.255 S\\ 0.21 S\\ 0.10  S\end{tabular}   \\ \hline
NGC 1792 (b)  & \begin{tabular}[c]{@{}c@{}} 0.2 \\ 1.07 \end{tabular} & \begin{tabular}[c]{@{}c@{}} -0.225\\ 0.261 \end{tabular} & \begin{tabular}[c]{@{}c@{}}0.37\\ 1.56\end{tabular}                 & \begin{tabular}[c]{@{}c@{}}0.026 S\\ 0.042 S\end{tabular}                 & \begin{tabular}[c]{@{}c@{}}1.07\\ 1.96\\ 2.84\end{tabular} & \begin{tabular}[c]{@{}c@{}}0.25 S\\ 0.24 S\\ 0.20 S\end{tabular}\\ \hline
NGC 2903 (a)  & \begin{tabular}[c]{@{}c@{}}0.387\\ 0.596 \end{tabular}   & \begin{tabular}[c]{@{}c@{}}0.444\\ -0.022\end{tabular}                         & \begin{tabular}[c]{@{}c@{}}0.13\\ 0.53\end{tabular} & \begin{tabular}[c]{@{}c@{}}0.135 R\\ 0.024 S\end{tabular}    & \begin{tabular}[c]{@{}c@{}} 0.324\\ 0.634\end{tabular}              & \begin{tabular}[c]{@{}c@{}} 0.177  R\\ 0.141 S\end{tabular}              \\ \hline
NGC 2903 (b)  & \begin{tabular}[c]{@{}c@{}}0.025\\ 0.314\end{tabular}  & \begin{tabular}[c]{@{}c@{}} -0.892\\ 0.578\end{tabular} & 0.12  & 0.116 S & \begin{tabular}[c]{@{}c@{}}0.093\\ 0.263\end{tabular}  & \begin{tabular}[c]{@{}c@{}}0.470 S\\ 0.167 S\end{tabular} \\ \hline
NGC 2903 (c)  & \begin{tabular}[c]{@{}c@{}}0.359\\ 1.62\end{tabular}   & \begin{tabular}[c]{@{}c@{}}0.336\\ -0.077\end{tabular}    & \begin{tabular}[c]{@{}c@{}} 0.424\\ 1.71\end{tabular} & \begin{tabular}[c]{@{}c@{}} 0.051 R\\ 0.009 S\end{tabular}  & \begin{tabular}[c]{@{}c@{}}0.53\\ 1.55\end{tabular}                        & \begin{tabular}[c]{@{}c@{}}0.091 R\\ 0.059  S\end{tabular}     \\ \hline
NGC 2903 (d)  & \begin{tabular}[c]{@{}c@{}} 1.32\\ 2.44\end{tabular}   & \begin{tabular}[c]{@{}c@{}} 0.86\\ -1.\end{tabular}   & \begin{tabular}[c]{@{}c@{}}0.42\\  2.32\end{tabular}          & \begin{tabular}[c]{@{}c@{}}1.11 S\\  0.057 B\end{tabular} & \begin{tabular}[c]{@{}c@{}}0.42\\ 2.32\end{tabular}         & \begin{tabular}[c]{@{}c@{}}0.505  S\\ 0.257 B\end{tabular}    \\ \hline
NGC 3059      & \begin{tabular}[c]{@{}c@{}} 0.435\\ 0.932\end{tabular}  & \begin{tabular}[c]{@{}c@{}}0.592\\ -1.\end{tabular}   & 0.26   & 0.33 B & \begin{tabular}[c]{@{}c@{}} 0.24\\ 0.74 \end{tabular} & \begin{tabular}[c]{@{}c@{}} 0.928 B\\ 0.53 B\end{tabular}                  \\ \hline
NGC 4303 (a)  & \begin{tabular}[c]{@{}c@{}}0.037 \\ 0.96\end{tabular} & \begin{tabular}[c]{@{}c@{}} 0.134 \\-0.835\end{tabular} & \begin{tabular}[c]{@{}c@{}}0.09\\ 1.72\end{tabular}     & \begin{tabular}[c]{@{}c@{}}0.033 R\\ 0.1  B\end{tabular}   & \begin{tabular}[c]{@{}c@{}} 0.15\\ 1.69\end{tabular}  & \begin{tabular}[c]{@{}c@{}} 0.091 R\\ 0.217 B\end{tabular}   \\ \hline
NGC 4303 (b)  & \begin{tabular}[c]{@{}c@{}} 0.26\\ 0.67\end{tabular}  & \begin{tabular}[c]{@{}c@{}} 0.77\\ -1.\end{tabular}   &  1.92  &  0.18 B    & 1.93 &  0.46 B  \\ \hline
NGC 4321 (a)  & \begin{tabular}[c]{@{}c@{}}0.57\\ 1.19\end{tabular} & \begin{tabular}[c]{@{}c@{}} 0.168\\ -0.273\end{tabular}     & \begin{tabular}[c]{@{}c@{}}0.303\\ 0.605\end{tabular}    & \begin{tabular}[c]{@{}c@{}}0.086 R \\ 0.084 B\end{tabular}           & \begin{tabular}[c]{@{}c@{}}0.31\\0.67\end{tabular} & \begin{tabular}[c]{@{}c@{}}0.267 R\\ 0.18 B\end{tabular}       \\ \hline
NGC 4321 (b)  & \begin{tabular}[c]{@{}c@{}} 0.61\\ 1.12\end{tabular}   & \begin{tabular}[c]{@{}c@{}} 0.24\\ -0.35\end{tabular}    & \begin{tabular}[c]{@{}c@{}}0.65\\ 3.46\end{tabular}                 & \begin{tabular}[c]{@{}c@{}}0.103 B\\0.054 B\end{tabular}   & \begin{tabular}[c]{@{}c@{}}0.68\\ 4.05\end{tabular}    & \begin{tabular}[c]{@{}c@{}}0.194 B\\ 0.08 B\end{tabular}     \\ \hline
NGC 4536 (a)  &  \begin{tabular}[c]{@{}c@{}}0.23\\ 1.29\end{tabular}  & \begin{tabular}[c]{@{}c@{}}0.03\\ -0.874\end{tabular}   & 0.252    & 0.201 B  & 1.07 & 0.118 B  \\ \hline
NGC 4536 (b)  & \begin{tabular}[c]{@{}c@{}}1.21\\ 2.61\end{tabular}     & \begin{tabular}[c]{@{}c@{}}-1.\\ 0.459\end{tabular}   &  2.49 & 0.145 B  &  2.56   &  0.325 B \\ \hline
NGC 4689 (a)  & \begin{tabular}[c]{@{}c@{}} 0.08\\0.25\end{tabular} &\begin{tabular}[c]{@{}c@{}} 0.47\\ -0.148\end{tabular}  & 0.204    & 0.049 R  & 0.63  &  0.150 S   \\ \hline
NGC 4689 (b)  & \begin{tabular}[c]{@{}c@{}}0.1\\1.04\end{tabular} &\begin{tabular}[c]{@{}c@{}}0.5\\ -0.112\end{tabular}  & \begin{tabular}[c]{@{}c@{}}0.298\\ 1.33\end{tabular}  & \begin{tabular}[c]{@{}c@{}}0.039  R \\ 0.02  S\end{tabular}  & \begin{tabular}[c]{@{}c@{}}0.556\\ 1.75\end{tabular} & \begin{tabular}[c]{@{}c@{}}0.135 R \\ 0.66 S\end{tabular} \\ \hline
NGC 5248 (a)  &  \begin{tabular}[c]{@{}c@{}}0.369\\0.62\end{tabular}  &\begin{tabular}[c]{@{}c@{}} -0.264\\ 0.04\end{tabular}  & 0.124    & 0.071 R  & \begin{tabular}[c]{@{}c@{}}0.17\\ 0.465\end{tabular} & \begin{tabular}[c]{@{}c@{}}0.325 R\\ 0.204 B\end{tabular}  \\ \hline
NGC 5248 (b)  & \begin{tabular}[c]{@{}c@{}} 0.03\\ 0.193\end{tabular}  & \begin{tabular}[c]{@{}c@{}} -1.\\ 0.791\end{tabular}   & 0.11  & 0.077 R  & \begin{tabular}[c]{@{}c@{}}0.164\\ 0.41\end{tabular}   & \begin{tabular}[c]{@{}c@{}}0.325 B\\ 0.208 B\end{tabular} \\ \hline
NGC 5248 (c)  & \begin{tabular}[c]{@{}c@{}} 1.87\\ 3.08\end{tabular}  & \begin{tabular}[c]{@{}c@{}} -0.681\\ 0.322\end{tabular}  & 0.113  & 0.071 R & \begin{tabular}[c]{@{}c@{}}0.155\\ 0.423\end{tabular}   & \begin{tabular}[c]{@{}c@{}}0.325 R\\ 0.204 B\end{tabular}                         \\ \hline
NGC 5921      & \begin{tabular}[c]{@{}c@{}}0.212\\ 0.511\end{tabular}    & \begin{tabular}[c]{@{}c@{}}0.39\\ -0.701\end{tabular}  & 0.13  & 0.090    B & 0.13   & 0.18 B  \\ \hline
\end{tabular}%
}\\
{{\textit{Notes.} The letters B, S and R are indication for Bars, Spirals and Rings. The corresponding set of observations for each galaxy is indicated following the same pattern as in Fig. \ref{NGC1300_m0_m1} to \ref{NGC5921_m0_m1}. }}
\end{table*}

\begin{figure*}
\begin{center}

  \includegraphics[scale=0.21]{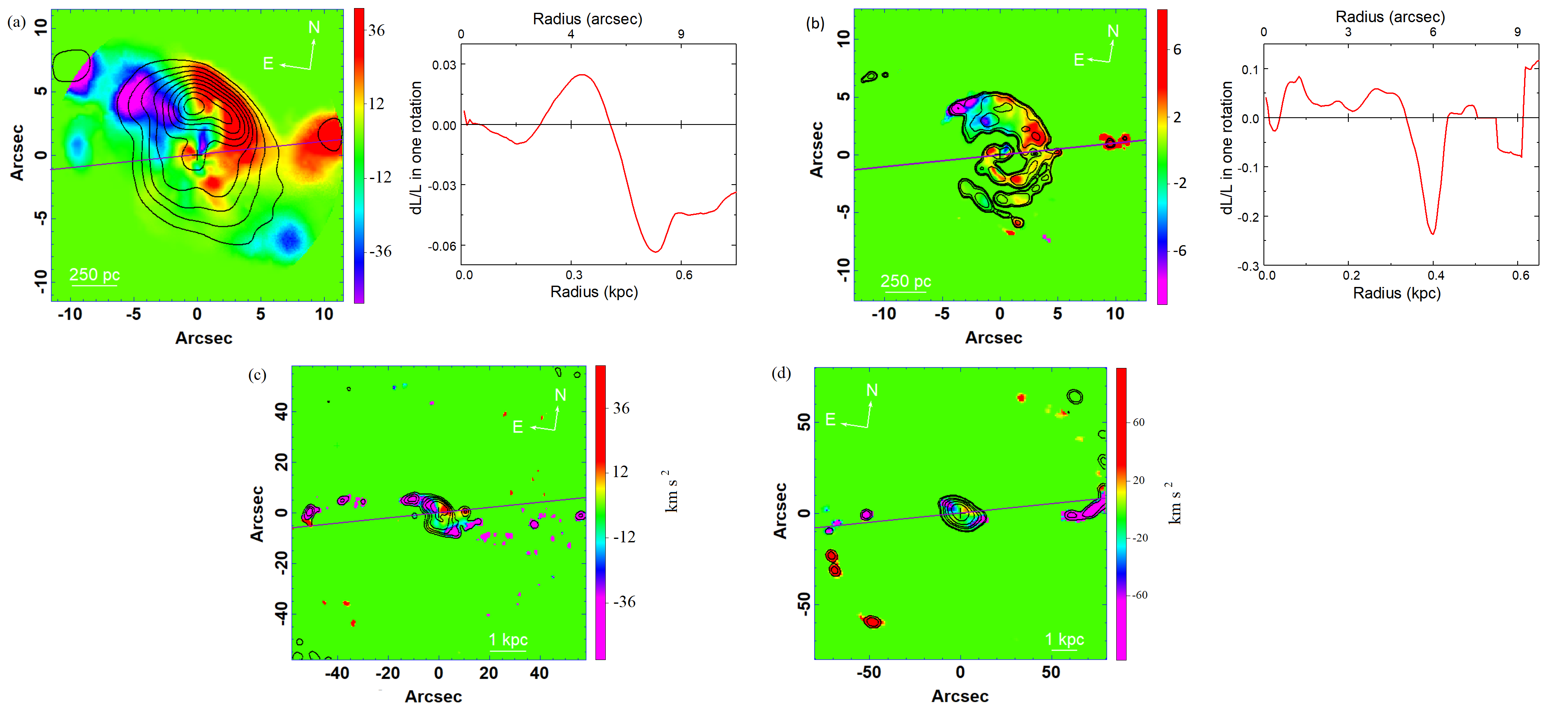}
  \caption{- NGC 1300 torque maps from the same set of data as in panel (a) Fig.~\ref{NGC1300_m0_m1}(a), (b) Fig.~\ref{NGC1300_m0_m1}(b), (c) Fig.~\ref{NGC1300_m0_m1}(c), and (d) Fig.~\ref{NGC1300_m0_m1}(d). The images were rotated and de-projected. The orientation N-E is indicated in the torque maps. The crosses represent the centre of the galaxy estimated from the emission peaks of the infrared images and ALMA data. The dark violet line represents the PA of the bar after rotation. Each map is associated with a graph that represents the radial distribution of the torque, averaged over the azimuth, and normalised to the angular momentum, per rotation at this radius. The colour scales in the maps are in arbitrary units (except for panels c and d), of the torques t(x,y) weighted by the gas surface density $\Sigma$(x,y). The contours of $\Sigma$(x,y) are from the mean molecular gas map (a) one to nine times 18.55 Jy/beam km/s, (b) 1, 1.1, 1.4, 2.5, 5.9, 16.5 times 0.3 Jy/beam km/s (c) 1, 5, 11, 23.8, 51.5 times 1.65 Jy/beam km/s, and (d) 1, 2.5, 12.5, 27.2, 58.9, 127.2 times 0.49 Jy/beam km/s. The velocities are relative to the centre.  \label{torquemaps_N1300}}
  
\end{center}
\end{figure*}

\begin{figure*}
\begin{center}

  \includegraphics[scale=0.21]{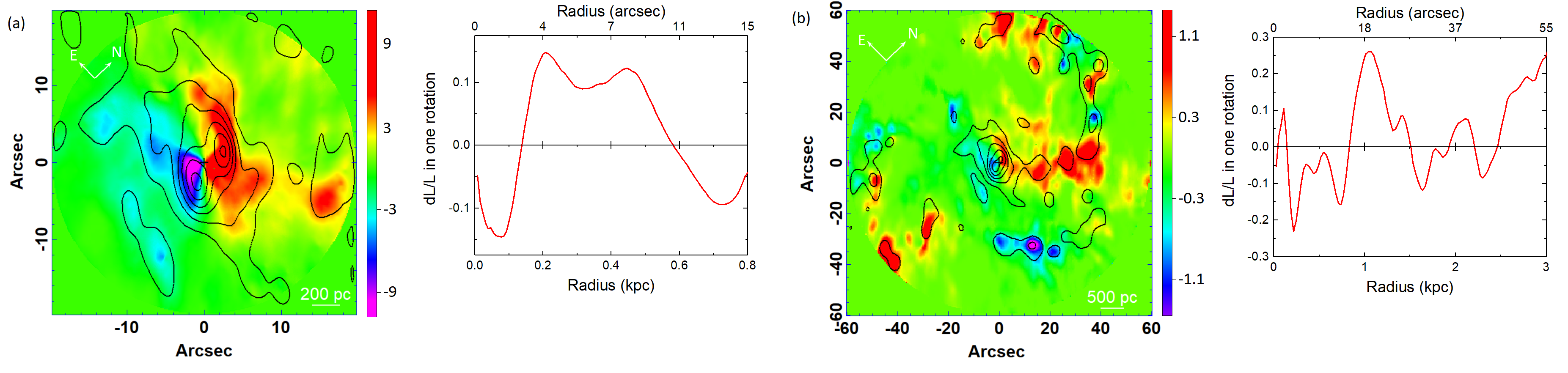}
  \caption{- NGC 1792 torque maps from the same set of data as in panel (a) Fig.~\ref{NGC1792_m0_m1}(a) and (b) Fig.~\ref{NGC1792_m0_m1}(b). Same caption as Fig.\ref{torquemaps_N1300}. The contour levels were taken from the mean molecular gas map (a) one to six times 34.13 Jy/beam km/s and (b) one to five times 32.11 Jy/beam km/s. \label{torquemaps_N1792}}
  
\end{center}
\end{figure*}

\begin{figure*}
\begin{center}

  \includegraphics[scale=0.21]{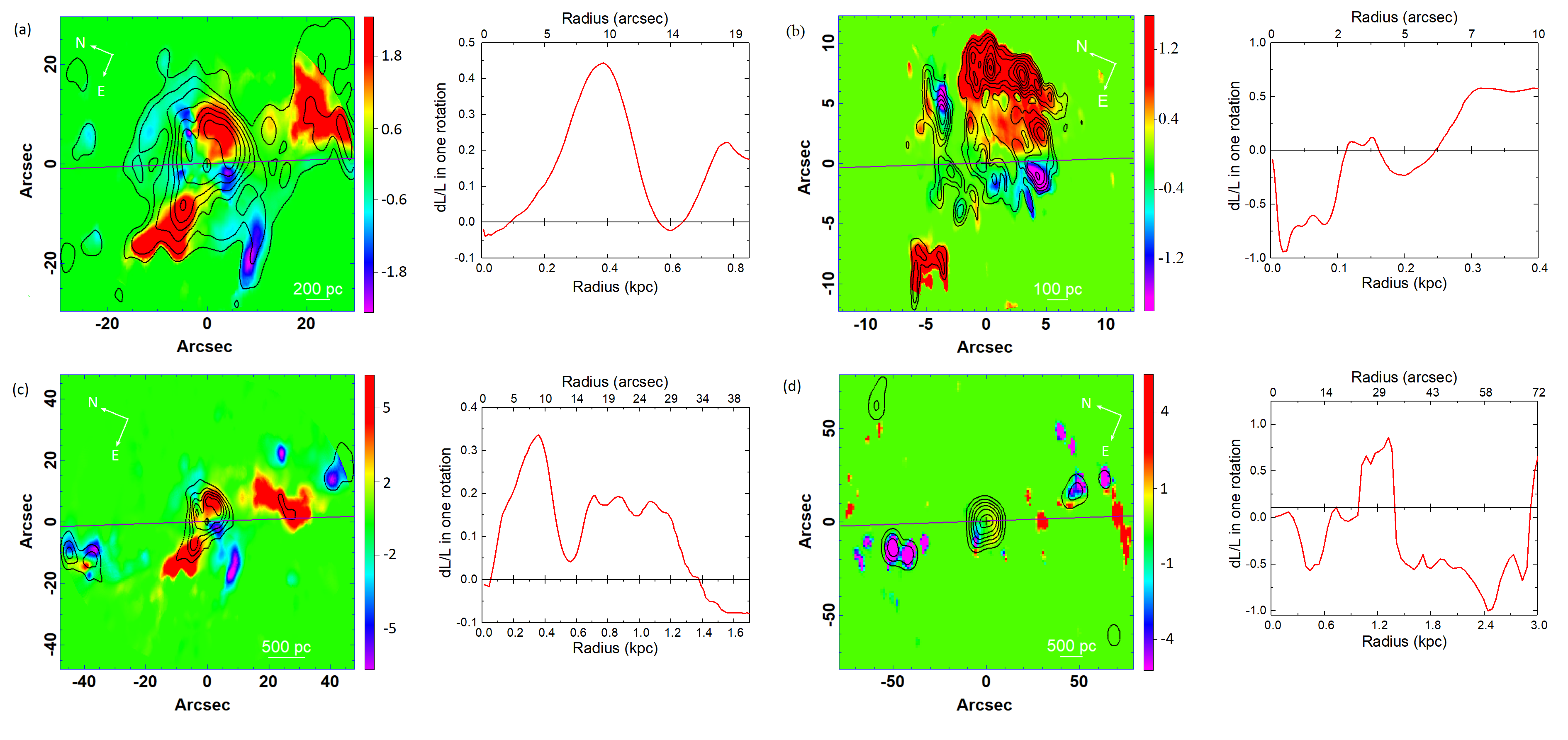}
  \caption{- NGC 2903 torque maps from the same set of data as in panel (a) Fig.~\ref{NGC2903_m0_m1}(a), (b) Fig.~\ref{NGC2903_m0_m1}(b), (c) Fig.~\ref{NGC2903_m0_m1}(c), and (d) Fig.~\ref{NGC2903_m0_m1}(d). Same caption as Fig.\ref{torquemaps_N1300}. The contour levels were taken from the mean molecular gas map (a) 1, 4, 9, 16, 25, 36, and 49, times 6.8 Jy/beam km/s, (b) one to seven times 1.08 Jy/beam km/s, (c) one to five times 1.43 Jy/beam km/s, and (d) 1, 4, 9, 16, 25, and 36 times 6.3 Jy/beam km/s.  \label{torquemaps_N2903}}
  
\end{center}
\end{figure*}

\begin{figure}
\begin{center}

  \includegraphics[scale=0.2]{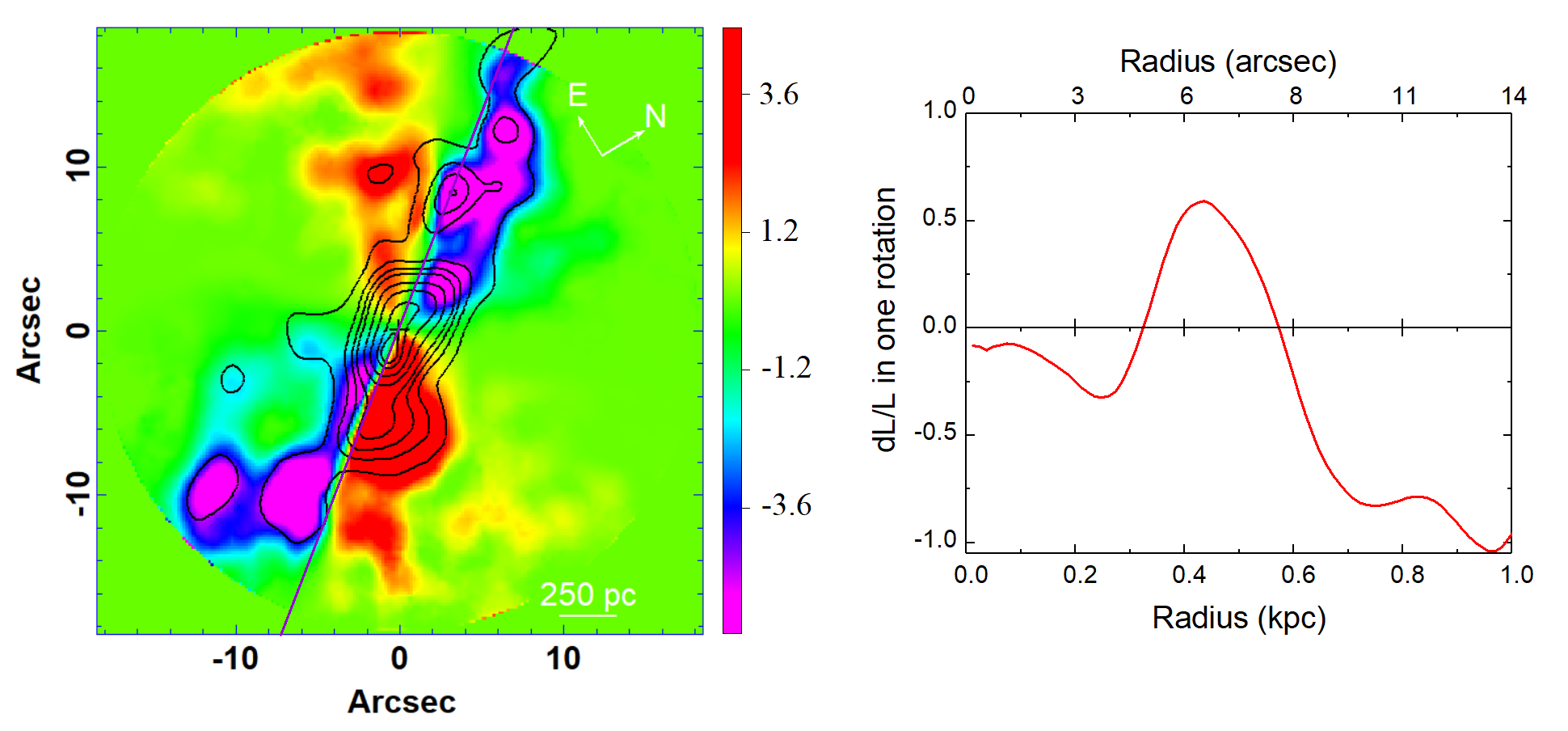}
  \caption{- NGC 3059 torque maps from the same set of data as in Fig.~\ref{NGC3059_m0_m1}. Same caption as Fig.\ref{torquemaps_N1300}. The contour levels were taken from the mean molecular gas map 1, 2, 3, 4, 6, 8, 10, and 11 times 15.7 Jy/beam km/s.   \label{torquemaps_N3059}}
  
\end{center}
\end{figure}

\begin{figure*}
\begin{center}

  \includegraphics[scale=0.21]{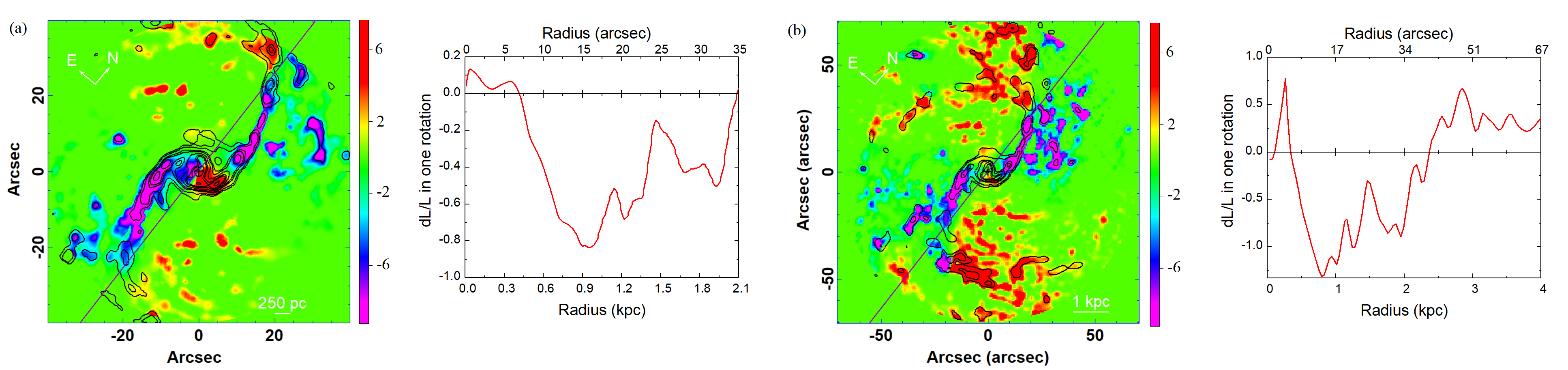}
  \caption{- NGC 4303 torque maps from the same set of data as in panel (a) Fig.~\ref{NGC4303_m0_m1}(a) and (b)Fig.~\ref{NGC4303_m0_m1}(b). Same caption as Fig.\ref{torquemaps_N1300}. The contour levels were taken from the mean molecular gas map (a) 1, 2, 4, 5, 7, 9, 10, and 12 times 14.6 Jy/beam km/s, and (b) 1, 2, 5, 6, 8, 9, and 10 times 15.9 Jy/beam km/s.   \label{torquemaps_N4303}}
  
\end{center}
\end{figure*}

\begin{figure*}
\begin{center}

  \includegraphics[scale=0.22]{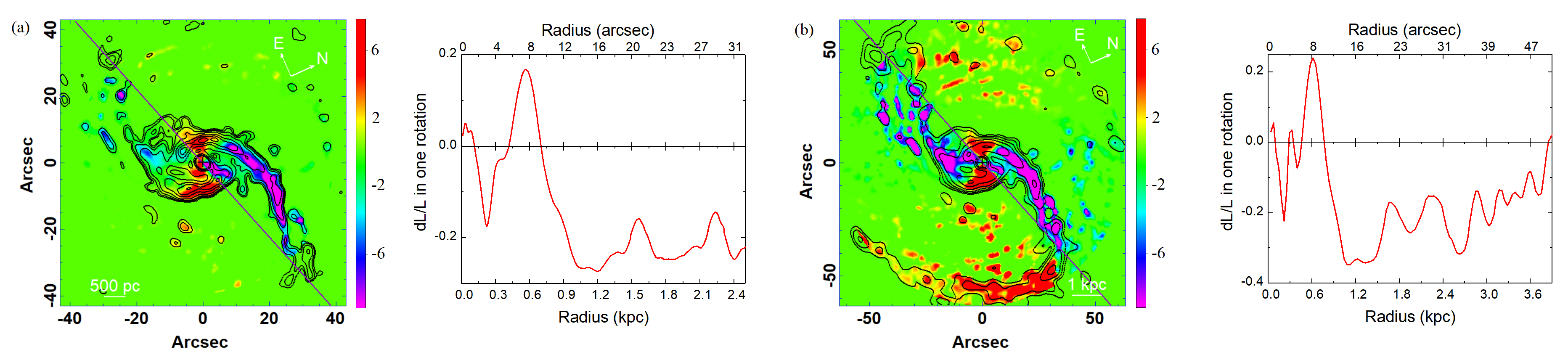}
  \caption{- NGC 4321 torque maps from the same set of data as in panel (a) Fig.~\ref{NGC4321_m0_m1}(a) and (b)Fig.~\ref{NGC4321_m0_m1}(b). Same caption as Fig.\ref{torquemaps_N1300}.  The contour levels were taken from the mean molecular gas map (a and b) 1, 2, 5, 10, 23, 34, and46 times 4.4 Jy/beam km/s.   \label{torquemaps_N4321}}
  
\end{center}
\end{figure*}

\begin{figure*}
\begin{center}

  \includegraphics[scale=0.22]{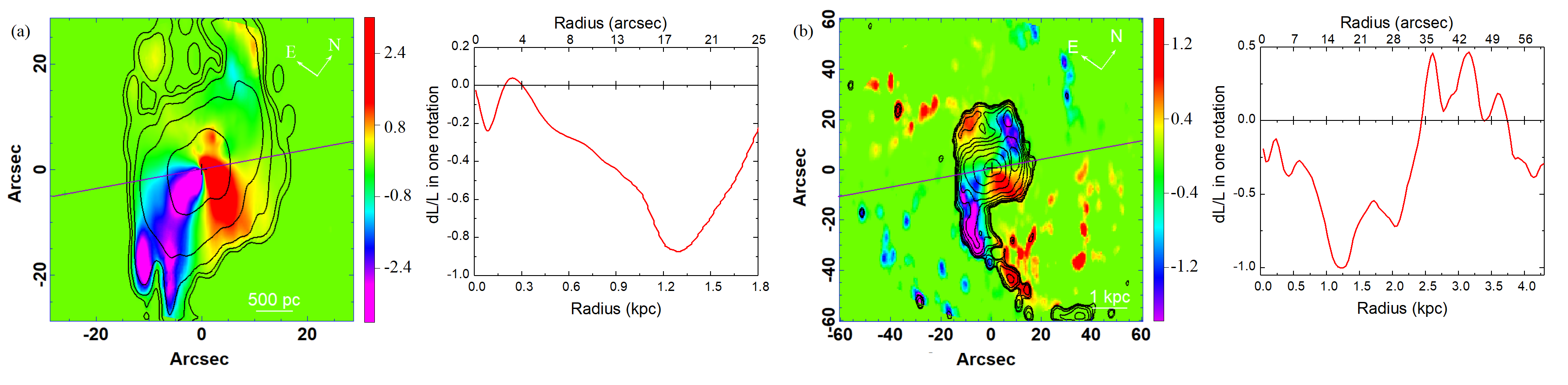}
  \caption{- NGC 4536 torque maps from the same set of data as in panel (a) Fig.~\ref{NGC4536_m0_m1}(a) and (b) Fig.~\ref{NGC4536_m0_m1}(b). Same caption as Fig.\ref{torquemaps_N1300}.  The contour levels were taken from the mean molecular gas map (a) 1, 4.2, 14.2, 45.8, and 145.8 times 1.56 Jy/beam km/s, and (b) 1, 1.1, 1.4, 2, 3.1, 5.5, 10, 19.3, 37.6, and 74.1 times 5 Jy/beam km/s.  \label{torquemaps_N4536}}
  
\end{center}
\end{figure*}

\begin{figure*}
\begin{center}

  \includegraphics[scale=0.21]{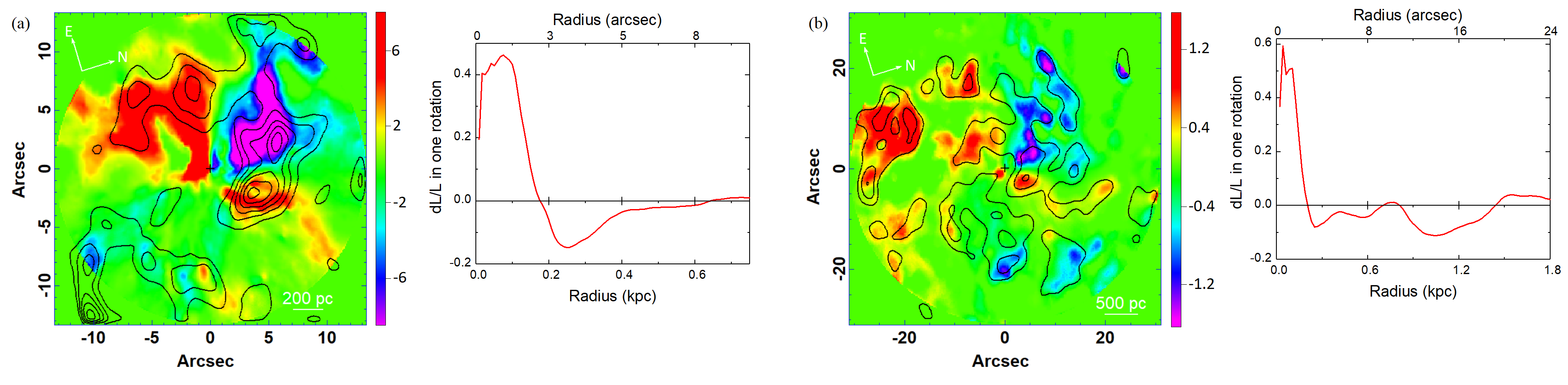}
  \caption{- NGC 4689 torque maps from the same set of data as in panel (a) Fig.~\ref{NGC4689_m0_m1}(a) and (b)Fig.~\ref{NGC4689_m0_m1}(b). Same caption as Fig.\ref{torquemaps_N1300}. The contour levels were taken from the mean molecular gas map (a) 1, 1.4, 1.7, 2, 2.2, and 2.5 times 6.8 Jy/beam km/s, and (b) one to three times 4.9 Jy/beam km/s.  \label{torquemaps_N4689}}
  
\end{center}
\end{figure*}

\begin{figure*}
\begin{center}

  \includegraphics[scale=0.21]{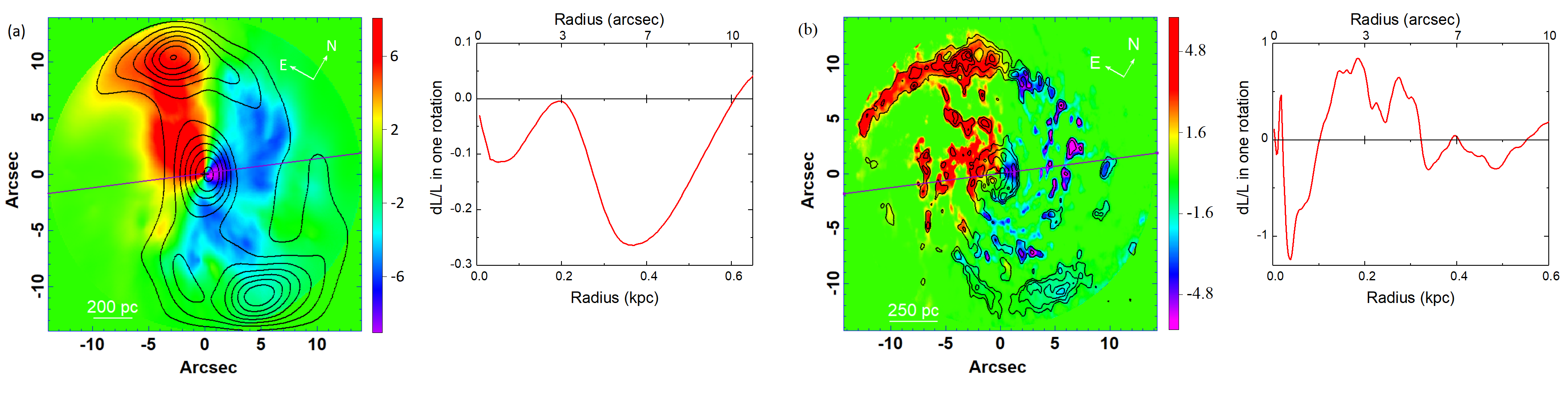}
  \caption{- NGC 5248 torque maps from the same set of data as in panel (a) Fig.~\ref{NGC5248_m0_m1}(a), (b)Fig.~\ref{NGC5248_m0_m1}(b), and (c)Fig.~\ref{NGC5248_m0_m1}(c). Same caption as Fig.\ref{torquemaps_N1300}. The contour levels were taken from the mean molecular gas map (a) 1, 1.4, 1.7, 2, 2.2, 2.5, and 2.7 times 59.3 Jy/beam km/s, and (b) one to five times 0.25 Jy/beam km/s.}  \label{torquemaps_N5248}

\end{center}
\end{figure*}

\begin{figure}
\begin{center}

  \includegraphics[scale=0.21]{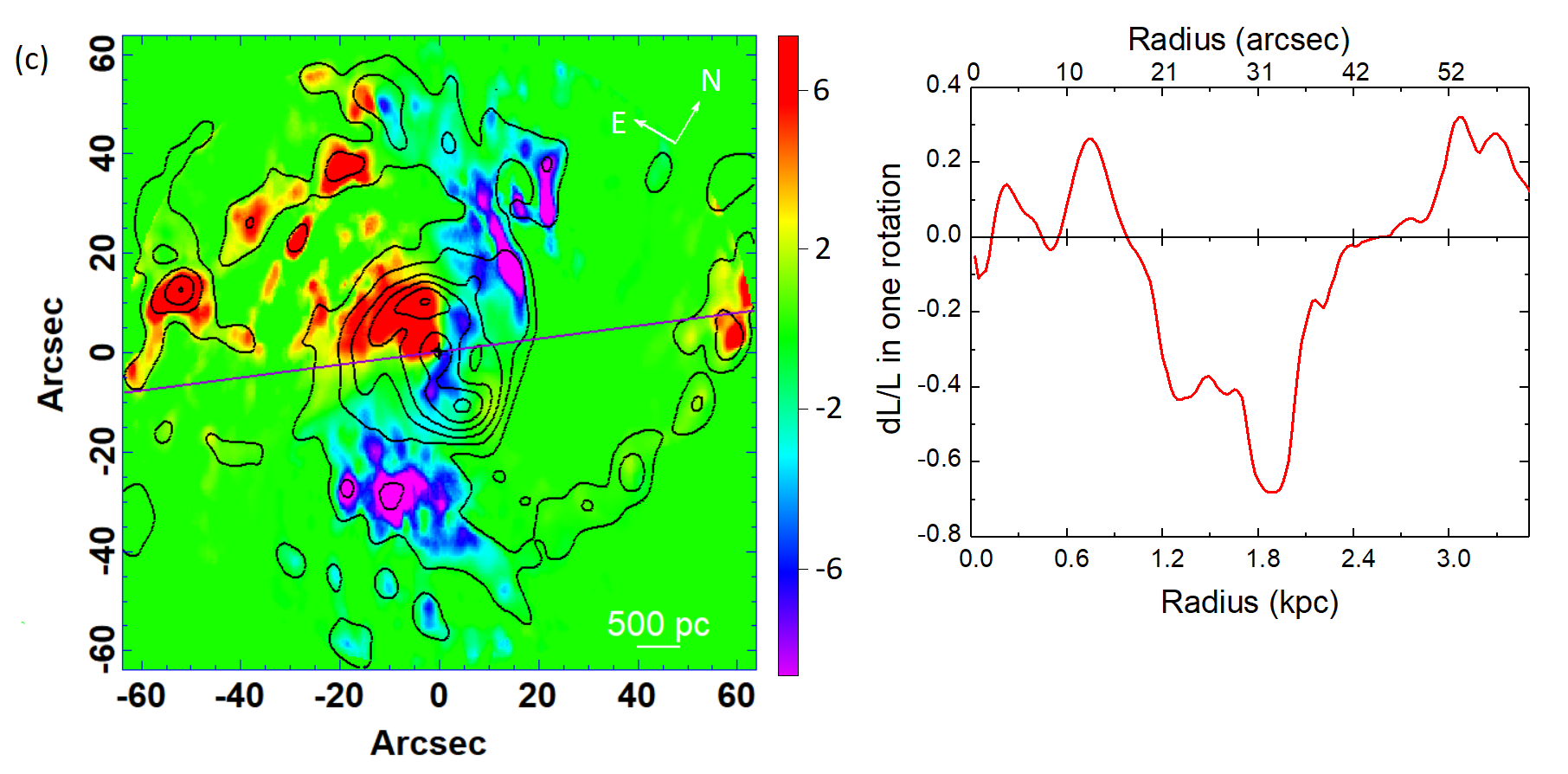}
  \caption{- cont. of Fig.~\ref{torquemaps_N5248}. The contour levels were taken from the mean molecular gas map (c) 1, 4, 9, 16, 25, and 36 times 3.48 Jy/beam km/s.  }  \label{torquemaps_N5248b}

\end{center}
\end{figure}

\begin{figure}
\begin{center}

  \includegraphics[scale=0.21]{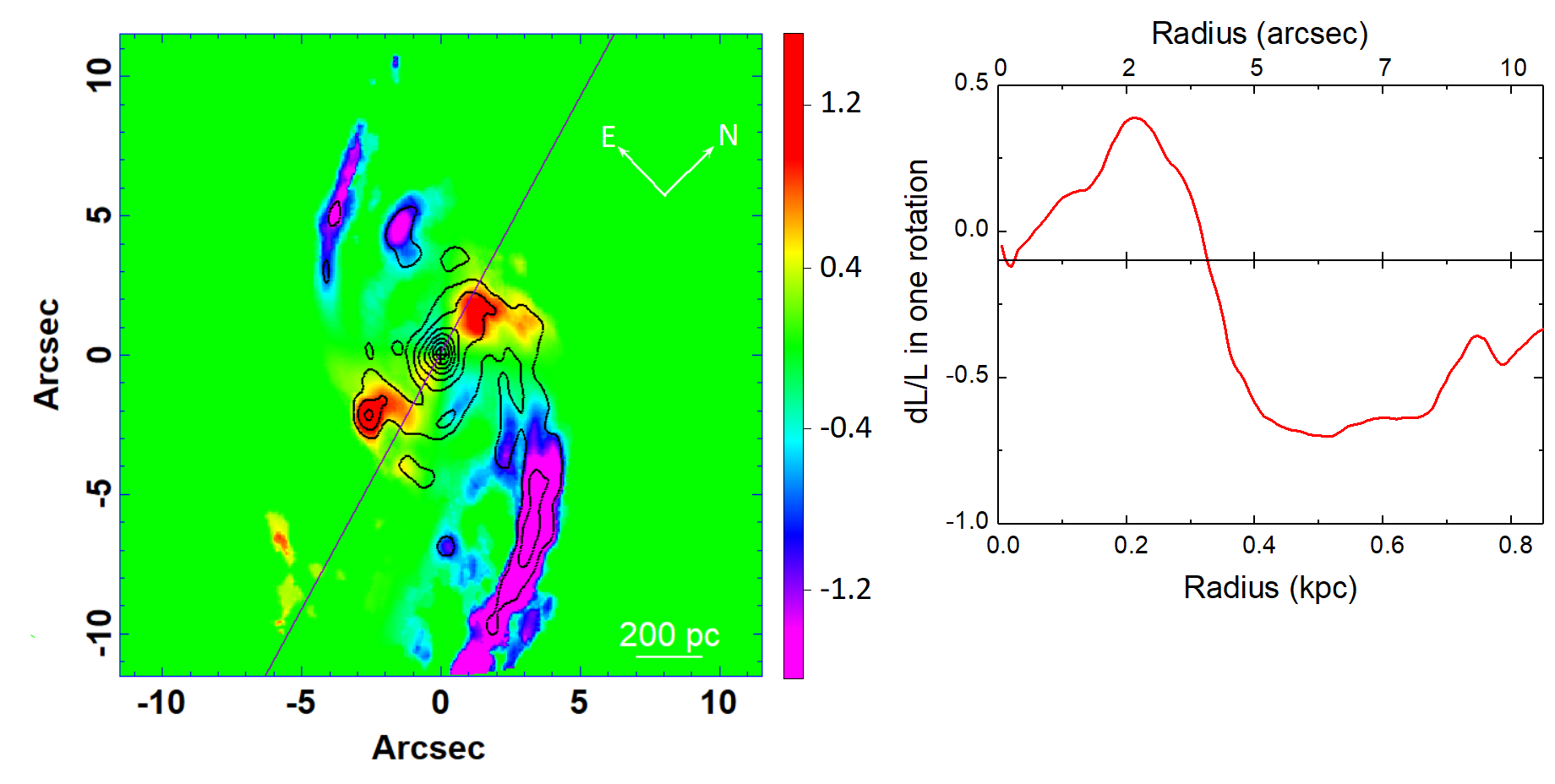}
  \caption{- NGC 5921 torque maps from the same set of data as in Fig.~\ref{NGC5921_m0_m1}. Same caption as Fig.\ref{torquemaps_N1300}. The contour levels were taken from the mean molecular gas map one to seven times 0.74 Jy/beam km/s.  }  \label{torquemaps_N5921}

\end{center}
\end{figure}

\subsection{Gas torque analysis}
\label{torq-analysis}

  The computation of the gravity potential from the deprojected red images 
  yields the estimate of the bar and spiral strength as a function of radius.
  The nature of the pattern (bar or spiral) is indicated by the phase
  and its variation with radius of the different Fourier components into which 
the potential $\Phi(R,\theta)$ is decomposed,
  $$
  \Phi(R,\theta) = \Phi_0(R) + \sum_m \Phi_m(R) \cos (m \theta - \phi_m(R)).
  $$
  \noindent
  The strength of the $m$-Fourier component, $Q_m(R)$, is defined as
  $Q_m(R)=m \Phi_m / R | F_0(R) |$, that is, by the ratio between tangential
  and radial forces \citep{Combes1981}.
  The strength of the total non-axisymmetric perturbation $Q_T(R)$ is defined
similarly, with the maximum amplitude of the tangential force $F_T^{max}(R)$.
  Their radial distributions and the radial phase variations are displayed 
  for a typical galaxy in Fig.~\ref{fig:qplot}. We determine in Table~\ref{tab:qplot}
  the radii and maximum amplitudes of $Q_T(R)$ over the disk for all galaxies.

  For each galaxy, the forces per unit mass ($F_x$ and $F_y$) at
  each pixel were obtained by deriving the potential. 
  The torques per unit mass $t(x,y)$ were then computed by
  $$
  t(x,y) = x~F_y -y~F_x.
  $$
  The sign of the torque was determined relative to the sense of rotation in the plane of the galaxy.
  The net effect on the gas at each radius is then 
  the product of the torque by the gas density $\Sigma$.  This quantity, $t(x,y)\times
  \Sigma(x,y)$, is shown in the left panels of Figs.~\ref{torquemaps_N1300}~--~\ref{torquemaps_N5921}.

  The torque weighted by the gas density $\Sigma(x,y)$ was then averaged over azimuth, 
  $$
  t(R) = \frac{\int_\theta \Sigma(x,y)\times(x~F_y -y~F_x)}{\int_\theta \Sigma(x,y)}.
  $$
\noindent
  The quantity $t(R)$ represents the time derivative of the specific angular momentum 
  $L=\Omega R^2$ of the gas averaged
  azimuthally \citep{Garcia-Burillo2005}. The efficiency of the gas flow was then estimated
  through normalising $t(R)$ at each radius by the angular momentum $L$, divided by
  the rotation period $T= 2\pi/\Omega$, as shown in the right panels of Figs.~\ref{torquemaps_N1300}~--~\ref{torquemaps_N5921}. With this normalisation, the computations are independent of the constant M/L ratios used.

  These figures show some general features. Negative torques, implying
  gas infall between corotation and ILR, as expected, are indeed observed in several barred galaxies
  when the FOV is sufficient and includes enough of the bar region. This is the case for
  NGC~1300 in Figs.~\ref{torquemaps_N1300} (c) and (d), and also NGC~4303, NGC~4321, NGC~4536, NGC~5248, and
  NGC~5921. When only the nuclear ring is included in the FOV, the situation is more complex,
  and the torques do not show a single coherent behaviour, as for NGC~1300 in 
  Figs.~\ref{torquemaps_N1300} (a) and (b).
  The torques are not significant in non-barred flocculent
  galaxies, as for NGC~1792 in Fig.~\ref{torquemaps_N1792} and
  NGC~4689 in Fig.~\ref{torquemaps_N4689}.

  NGC~2903 is a special case (Fig.~\ref{torquemaps_N2903}). Although barred, with a coherent spiral structure, 
  the torque values are more chaotic. This is due to several factors. The main factor is the
  patchiness of the CO emission, which prevents a spatial continuity in the torque computation.
   We note that the torques we computed from the two different fields of view are
   sometimes different. The reason is the different gas maps that were obtained with different configurations
   of the ALMA interferometer, with different spatial resolutions.
  Another reason is that the red image of NGC~2903 is quite dusty, with conspicuous dust lanes. This
  leads to an irregular potential estimation that differs from the real potential.

   NGC~3059 has a very thin bar, almost on the minor axis. It is therefore even longer
   when deprojected, although the inclination is quite low. The torque is positive from 0.3 to 0.6~kpc,
   and then negative from 0.6 to 1~kpc (Fig~\ref{torquemaps_N3059}). This would mean that gas is driven to a ring at a
   radius of 0.6~kpc. However, in this case, the gas orbits are expected to be very elongated, and 
 positive and negative torques might partially cancel out over the
   orbits. The torques are negative at radii lower than 0.3~kpc, which explains the gas
   concentration inside a radius of 3 arcsec, that is, 220~pc.

\begin{figure}
\begin{center}

  \includegraphics[scale=0.33]{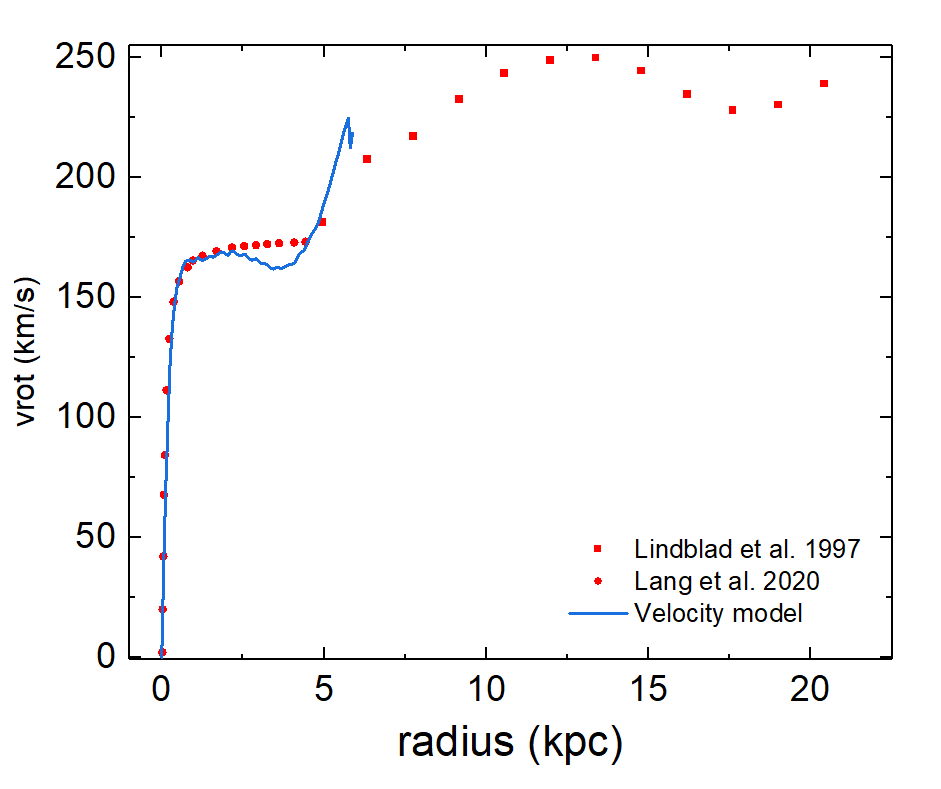}
  \caption{-Gas velocity rotation curves from HI \citep{Lindblad97} (red squares) and from CO \citep{Lang20} (red points) of NGC 1300 used to compute the torque map values of the data sets in panels (c) and (d). Only the highest points of each curve were kept. The blue curve represents the velocity model obtained from the computed potential (with limitations of the red image cut-off beyond 5~kpc).\label{fig:Vrot-1300}}
  
\end{center}
\end{figure}

  The quantification of the torques for each galaxy and each FOV is displayed in
  Table~\ref{tab:qplot}. The relative loss and gain of angular momentum per rotation is 
  higher than one in absolute value only for NGC 1300, and it is not indicated in this table.
  It does not satisfy the main hypothesis
  underlying the azimuthal average of the torque, that the gas has time to remain in its
  radial range during one rotation. In this case, we plot only the torque map, normalized 
  with physical quantities, computed with the rotation curve of the galaxy. The latter was derived from the potential itself, assuming a constant mass-to-light ratio, and
  normalised to the CO, HI, and optical observations (see Fig.~\ref{fig:Vrot-1300}). In this figure, 
  the rotation curve from the potential model is compared to the observed curves
  from \cite{Lang20}, \cite{england89}, and \cite{Lindblad97}.

\section{Discussion} \label{disc}

We studied ten MW analogues, and only two galaxies of this sample are considered non-barred. This sample is homogeneous when we consider that galactic bars may be in a 1-2 Gyr transient phase \citep[e.g.][]{Bournaud2002}
and when, as previous studies argued (see
Appendix \ref{revisiongalaxies}), the AGN activity is weak in these nuclei (which is also an intermittent characteristic, with a duty cycle of 40 Myr). The comparison of gas flow processes within this matched sample can help us to further understand the evolution of these types of galaxies, including our own. Although this sample is small because available data are limited, it is representative because it was selected based on objective criteria. The results are consistent for this sample. 

It is important to discuss the limitations of this work. One limitation is the spatial resolution of the data. One of the most frequently encountered questions in astronomy concerns the feeding and feedback processes in galactic nuclei. In order to study this matter comprehensively, high spatial resolution data are fundamental, because these processes occur in a very small region (10 -- 100~pc). The spatial resolution of the data we analysed here is insufficient to compute the gas dynamics at these scales in detail. As a consequence, we disregarded results below 1 arcsec (corresponding to 66~pc at an average distance of 13.7~Mpc) in general. The torque maps and values in this central region are
uncertain, not only because of the spatial resolution, but sometimes also due to phenomena associated with possible AGN, which may generate ambiguous results that can only be differentiated with data with a higher spatial resolution. These are rare. Not many galaxies are observed with the necessary instrumental configuration for an
analysis like this. The spatial resolution we used is considered moderate, and can only provide a general idea of what occurs in these galaxies at scales of 100~pc -- 1~kpc.

The assumption that dark matter has no effect is a good approach because these galaxies are massive enough (see the stellar masses in Table \ref{tabelagal}). In massive disk-dominated galaxies like this, dark matter is not dominant in the central region and shows a cored distribution \citep{deBlok2008}. In addition, we studied features in the nuclear regions of the objects, where baryons dominate the gravitational potential.

The torque maps were computed in each pixel, weighted by the gas density. To derive the global impact on the gas inflow/outflow, we averaged on the azimuth at each radius, although the orbits of the gas, when deprojected, are not circular. We then quantified the relative angular momentum variation in one circular rotation. Although gas orbits might be elliptical in these galaxies, this average gives the correct order of magnitude over many gas streamlines at a given radius. We did not considered this average when the relative loss of angular momentum was higher than one in absolute value (e.g. NGC~1300).

In the case of NGC 1300, we have two data sets (c and d), which are related to the large FOVs that provided noisier results, but also large variations in angular momentum. As a consequence, we could not take the self-normalised torque curves into account and had to use the velocity rotation curve (Fig. \ref{fig:Vrot-1300}) in order to normalise the
results. Some parts of the CO maps of these data sets show no emission, and the S/N in other parts is too low, which causes the torque curves to be noisy. The same was observed in other galaxies, but with fewer fluctuations.

Another limitation of this work is the determination of the PA and the galaxy inclination. This relies on the assumption that the outer optical isophotes are circular. These determinations can be biased in barred galaxies, where the PA of the bar can easily be misinterpreted to be aligned with the galaxy major axis, causing systematic errors. The inclination angle has the same problem because these galaxies are not
axisymmetric disks. The results strongly depend on these values because the deprojection is important for the computation of the torques.

We did not detect any dynamical influence of AGN feedback in these galaxies. One of the reasons, as mentioned, is the low spatial resolution and low AGN activity. We did not detect radio jets either. An optical data analysis (with a higher spatial resolution within the 500 pc scale) would be interesting in this case, in order to compare the ionized and molecular gas behaviour, and also to determine whether the AGN affects the surrounding ionized gas.

We provided the characteristics of gas flows in MW twins at 100~pc- 1~kpc scales. 
We consider this study as preliminary results that are to be pursued in the future with data that have a better spatial resolution, and perhaps with multiwavelength data. With these data, it will be possible to connect what we observe at these scales with the smaller scales and to quantify the AGN fueling mechanisms and/or the feedback on the gas flow in these galaxies. In ringed galaxies, the infall of gas is expected to be contained within the ring, but the gas can also be driven towards the nucleus through nuclear spirals/bars. 

\section{Conclusions} \label{conclusion}

We have quantified the gravity torques and their impact on the gas flows in a sample of eight MW analogues with two non-barred counterparts. The sample is homogeneous in mass, morphological type, and fundamental physical characteristics. The bar and spiral strengths varied, as did the AGN types, because they are considered variable during the lifetimes of typical spiral galaxies. This allowed us to compare the effect of non-axisymmetric features on the gas inflows and outflows. By using HST and Legacy images of the stellar emission and ALMA data of the molecular gas emission, we were able to compute the molecular gas torques in this sample at scales of hundreds of parsecs to kiloparsecs. 

The galaxy sample presented a diversification of structures, such as rings (NGC 1300, NGC 2903, NGC 4303, NCG 4321, NGC 5248, and NGC 5921) and spiral arms connected to them. The sample also contains flocculent galaxies (the best example is NGC 4689, without any recognisable structure).

Our multi-scale study showed that torques are not significant in the inner resonant rings, even when they are not circular. They are not significant for flocculent galaxies either. For barred galaxies, torques are negative between the corotation and the inner Lindblad resonance, as expected. The torques computed on the H$\alpha$ emitting gas are compatible with those on the molecular gas, but they are noisier and more patchy.

The relative loss of angular momentum can be quite high and may reach nearly 100\% in one rotation for extreme cases. When the bar is very strong and thin, the gas orbits are very elongated, which prevents the formation of a ring. The gas is then effectively driven to the nucleus. 

We studied intermediate scales between 100~pc and 1~kpc scales. The gas inflow
revealed by the torques produces starbursts in the centre of barred galaxies.
Future studies at higher spatial resolution are needed to follow-up the AGN feeding into the nucleus.

\begin{acknowledgements}
PS acknowledges Funda\c{c}\~ao de Amparo \`a Pesquisa do Estado de S\~ao Paulo (FAPESP) post-doctoral fellowships 2020/13239-5 and 2022/14382-1.
  The Legacy Surveys consist of three individual and complementary projects: the Dark Energy Camera Legacy Survey (DECaLS; Proposal ID \#2014B-0404; PIs: David Schlegel and Arjun Dey), the Beijing-Arizona Sky Survey (BASS; NOAO Prop. ID \#2015A-0801; PIs: Zhou Xu and Xiaohui Fan), and the Mayall z-band Legacy Survey (MzLS; Prop. ID \# 2016A-0453; PI: Arjun Dey). DECaLS, BASS and MzLS together include data obtained, respectively, at the Blanco telescope, Cerro Tololo Inter-American Observatory, NSF’s NOIRLab; the Bok telescope, Steward Observatory, University of Arizona; and the Mayall telescope, Kitt Peak National Observatory, NOIRLab. Pipeline processing and analyses of the data were supported by NOIRLab and the Lawrence Berkeley National Laboratory (LBNL). The Legacy Surveys project is honored to be permitted to conduct astronomical research on Iolkam Du’ag (Kitt Peak), a mountain with particular significance to the Tohono O’odham Nation. The Digitized Sky Surveys were produced at the Space Telescope Science Institute under U.S. Government grant NAG W-2166. The images of these surveys are based on photographic data obtained using the Oschin Schmidt Telescope on Palomar Mountain and the UK Schmidt Telescope. The plates were processed into the present compressed digital form with the permission of these institutions. This research is based on observations made with the NASA/ESA Hubble Space Telescope obtained from the Space Telescope Science Institute, which is operated by the Association of Universities for Research in Astronomy, Inc., under NASA contract NAS 5–26555. These observations are associated with program(s) 15133, 15654, 9788, 9042, 15323. This paper makes use of the following ALMA data: ADS/JAO.ALMA\#2015.1.00925.S, ADS/JAO.ALMA\#2015.1.00956.S, ADS/JAO.ALMA\#2017.1.00886.L, ADS/JAO.ALMA\#2018.1.00517.S, ADS/JAO.ALMA\#2018.1.01651.S, ADS/JAO.ALMA\#2019.1.01718.S, ADS/JAO.ALMA\#2019.1.01730.S, ADS/JAO.ALMA\#2019.1.00876.S, and ADS/JAO.ALMA\#2019.1.01742.S. ALMA is a partnership of ESO (representing its member states), NSF (USA) and NINS (Japan), together with NRC (Canada), NSTC and ASIAA (Taiwan), and KASI (Republic of Korea), in cooperation with the Republic of Chile. The Joint ALMA Observatory is operated by ESO, AUI/NRAO and NAOJ.
  This work was supported by the Programme National Cosmology et Galaxies (PNCG) of CNRS/INSU with INP and IN2P3, co-funded by CEA and CNES.
\end{acknowledgements}

%
\bibliographystyle{aa} 
\bibliography{bibliography} 
%

%

%

\begin{appendix} 

\section{Comments about individual galaxies:}\label{revisiongalaxies}

\textit{NGC 1300:} is a very well known MW morphological twin with a
prominent strong bar and two arms rich in star forming knots \citep{Eskridge02,Knapen02}.
Spiral arms starting at the end of the bar, extend out to the
outer Lindblad resonance \citep{england89}, responsible for the
outer pseudo-ring (major axis: 5'.71-- \citealt{buta93}, age: $10^9$ yrs --\citealt{elmegreen96}). 
The inner Lindblad resonance is the site of a circumnuclear
ring. Star formation is concentrated at the ends of
the bar \citep{elmegreen92}, and, in the central part, there is not much star formation,
no molecular gas concentration, and no clear HII regions \citep{Knapen02,maeda18}, possibly
due to the fast collision of gas clouds \citep{maeda18,maeda20,fujimoto20}. There are many
classifications for this galaxy nucleus. According to some works
there is no AGN (e. g. \citealt{martini03,stuber21}); while 
others see evidence of AGN activity (outflow: \citealt{saikia94}, low luminosity AGN: \citealt{jang14,constantin15,gadotti19,menezes22}). In any case, there is evidence of gas infall into the
inner galaxy (radius $<$ 25" \citealt{lindblad96}), which harbours an SMBH with  $log(M_{BH}/M_{\odot})= 7.82^{+0.29}_{−0.29}$ 
 \citep{atkinson05}. No evidence of
molecular gas outflow was detected according to \citet{stuber21}. The inner ring is
more continuous in the northern part, with distinct regions of star
formation and there is no evidence of a nuclear bar (\citealt{erwin04}, \citealt{perezramirez2000}, and references
therein). This ring appears complete in the NIR \citep{Knapen02} and its semi-major axis is
400 pc \citep{comeron10}, with age of $4.6 \pm 0.8$ Gyrs \citep{rosadobelza20}. The stellar velocity maps from MUSE
show rotation with the blueshift towards east and redshift towards west \citep{gadotti20}.

\textit{NGC 1792:} is classified as a non-barred SA(rs)bc galaxy \citep{RC3,Eskridge02,sun20,neumann23,stuber23}, with
flocculent \citep{Eskridge02} chaotic \citep{Garrison74} spiral arms rich in star-forming
regions \citep{dahlem94,Eskridge02}. The galaxy is
currently interacting with NGC 1808 \citep{dahlem92}. Its nuclear region, even though
showing [NeV] emission (that is commonly found in AGNs --\citealt{Goulding09,goulding10}, is
classified as non-active in many studies \citep{veron-cetty86,stuber21,menezes22,neumann23}.
\citet{Pereira-santaella10} say that
the [NeV] emission detection in this galaxy might be due to supernova
remnants, planetary nebulae, Wolf-Rayet stars or that its HII region
classification might be due to strong obscuration of the nuclear region
or to a very low luminosity AGN. The mass of the SMBH in its centre
was estimated as $log(M_{BH}/M_{\odot})= 6.83^{+0.12}_{−0.53}$  \citep{goulding10}. Observations at the H-band indicate the presence of a short nuclear bar \citep{jungwiert97}. There is no
evidence of molecular gas outflows \citep{stuber21}.

\textit{NGC 2903:} is a very well-studied SAB(rs)bc \citep{RC3} galaxy with a strong (in
the NIR --\citealt{regan99,Sheth02,hernandez05}, in short wavelengths it is similar to a patchy spiral disk --\citealt{popping10}) and young bar (\citealt{dutil99}, with age between 200 and 600 Myr -- \citealt{leon08}) immersed in a flocculent disk \citep{Canzian93}, somehow in a
very complex structure. Bright HII regions delineate the bar and the multi-arm structure of the
galaxy \citep{Knapen02}. This galaxy is a typical example of "hot-spot" galaxy \citep{sersic73,Turnrose76,bonatto89}, rich in individual young stellar clusters \citep{alonsoherrero01}, and starburst galaxy \citep{Turnrose76,wynn-williams85,Simons88,Jackson89,planesas97,perez-ramirez10}. The hot-spots near the nucleus seem part of the circumnuclear starburst ring \citep{planesas97,sakamoto99,perezramirez2000,alonsoherrero01,mazzuca11}, which has a radius of about 600 pc \citep{leitherer02}. The galaxy hosts a SMBH with a mass of $log(M_{BH}/M_{\odot})= 7.06^{+0.28}_{−7.06}$ \citep{vandenbosch16}, and is commonly
classified as non-active \citep{casuso90,ho97,sodre99,Pereira-santaella10,neumann23}, which is supported by no evidence of molecular gas outflows \citep{stuber21}, no significant [NeV] emission \citep{Goulding09}, and low luminosity in X-rays (0.5−8.0 keV --\citealt{yukita12}). However, NGC 2903 has NIR emission-line ratios compatible with an AGN \citep{matsuoka12} and new measurements of its X-ray luminosity indicate the presence of a variable AGN \citep{hoyer}. The nuclear starburst might be due to the bar that drives the molecular gas into the centre \citep{jackson91,hernandez05,leon08}.

\textit{NGC 3059:} is not a very well-studied galaxy. It is classified as SB(rs)bc; however some studies classify it as SBcd (e. g. \citealt{Eskridge02}), weakly barred \citep{stuber23} or with a small bar \citep{beck02}. There is no evidence
of an inner ring in the literature \citep{stuber23}, and no AGN \citep{stuber21,Goulding09,neumann23}.

\textit{NGC 4303:} is a grand design SAB(rs)bc \citep{RC3} double barred galaxy (\citealt{Colina97b,colina99,Colina2000}, nuclear bar diameter: 0.2 kpc, primary bar diameter: 2.64 kpc-- \citealt{erwin04}). It has a young (2.5 -25 Myrs) circumnuclear ring (200−250 pc-- \citealt{riffel16}), connected to the
bar (which is possibly responsible for the feeding of the nuclear
region and the ring-- \citealt{Colina97b}), by two spiral arms, and to an inner bar
\citep{Colina97b,Colina2000,perezramirez2000}. \citet{Schinnerer02} studied the molecular emission of the central kiloparsec of NGC 4303 and found that inflowing motions of the primary bar of the galaxy are responsible for the complex structure of the centre of this galaxy, and that this circumnuclear ring might be in an
early stage of formation. The nuclear activity of this galaxy was classified as Seyfert 2 or LINER or in-between (e. g. \citealt{colina99,Argyle90}), or HII region \citep{ho97}. According to \citet{colina02}, there is a young ($\sim$ 4
Myr) super star cluster coexisting with the low luminosity AGN in
the nucleus (within 3.1 pc), and \citet{Tzanavaris07,jimenez03} could not distinguish
its nuclear emission from a hidden AGN, a central stellar cluster,
or a ultra-luminous X-ray source. NGC 4303 harbours an SMBH with a
mass of $log (M_{BH}/M_{\odot})= 6.58^{+0.07}_{-0.26}$ \citep{davis19}. An
[OIII] outflow was detected north-east of the nucleus by \citet{colina99} and no molecular gas outflow was detected \citep{stuber21}.

\textit{NGC 4321:} is a very well studied MW morphological twin (SABbc-- \citealt{RC3}). Due to its inclination, distance, and contrasted features, it was subject to many detailed studies. For instance, this galaxy has a clumpy circumnuclear ring, whose peak radius is 458 pc \citep{song21}. Such starburst ring is being maintained by the gravitational influence of the bar \citep{arsenault88,knapen95b,perezramirez2000} and its age is about 50 Myrs (the majority of stars are younger than 6 Myr -- \citealt{sanchezgill11}). \citet{knapen95b} state that there are two Inner Lindblad Resonances in the nuclear region: one at 500 pc and the other at 1.3 kpc from the nucleus.
According to these authors, most of star formation happens at the inner resonance (also see \citealt{sakamoto95,yuan98,knapen2000}). The ring is connected by two arms leading the bar to the nucleus (\citealt{knapen95a,knapen95b,Knapen02} and the younger stellar populations are located near the connecting points between the ring and the larger bar \citep{mazzuca08}. There is also a double bar structure \citep{pierce86,shaw95}, the nuclear bar being clearly detected in the NIR and the primary bar being weaker \citep{font14} or moderate \citep{knapen95b}. Although \citet{knapen95b} do not found strong evidence for the second bar, they say both structures are part
of the same bar, aligned with comparable strength; \citet{garciaburillo98} dynamical models
favor that this nuclear bar is decoupled from the major bar and it is rotating faster. There are two dust arms connecting the nuclear bar to the outer bar \citep{knapen95b,sakamoto95}, and \citet{wada98} say that this nuclear structure is short lived (10$^7$ yrs) and recurrent. NGC 4321 nucleus was classified as HII region (\citealt{keel83}, supported by \citealt{gonzales09} that have not found any AGN activity); however, some studies classify it as a transition object \citep{ho97} and the emission of [NeV] detected by \citet{satyapal08}
sustains the hypothesis of a low luminosity AGN. The SMBH in NGC 4321 centre
has a mass of $log (M_{BH}/M_{\odot})= 6.67^{+0.17}_{-6.67}$ 6.67+0.17 −6.67 \citep{vandenbosch16}. The high nuclear CO concentration is a result of the gas being driven by gravitational torques from the bar; however the fact that there is no strong AGN activity at the centre of this galaxy shows that this inflow is still at intermediate scale, and is not yet prolonged into the center to trigger nuclear activity \citep{sakamoto95}. Such remark is actually
proven by the detailed torque analysis made by \citet{Garcia-Burillo2005}: there are positive torques along the spiral arms at the trailing edges of the nuclear bar and negative torques along the leading quadrants and outer disk (r $>$ 900). They also noticed that the average torque inside 200 pc is positive, implying that there is no significant inflow of molecular gas in the nucleus. \citet{castillo-morales07} detected a nuclear blueshifted region that might be related to an
outflow of gas. Such outflow was also reported by \citet{jimenezesvicente07} and the authors say that such wind was generated by a nuclear starburst.

\textit{NGC 4536:} is a SAB(rs)bc \citep{RC3} galaxy with two conspicuous  spiral arms \citep{mazzalay14}. This galaxy was previously classified as starburst \citep{ho97}; however \citet{hughes05} classified it as a weak AGN, based on optical emission-line ratios. There is [NeV] emission, detected 10" north-east from the optical nucleus \citep{satyapal08}, and XMM-Newton data presents a power law emission typical of low luminosity AGNs \citep{mcalpine11}, that supports the presence of a low luminosity AGN in this nucleus. A circumnuclear ring with radius of about 270 pc with age of 6.5 Myr was detected with near-infrared data, with no gradients along the ring \citep{mazzalay13,mazzalay14}. There is no evidence of molecular gas outflow \citep{stuber21} and, in the inner 500 pc, the gas kinematics is predominantly circular with streams of inflowing gas along bar into the centre \citep{jogee05}. The nucleus is connected to the circumnuclear ring and there is evidence of a nuclear bar \citep{mazzalay14}. The star formation is predominant in the ring and it has ceased in the nucleus \citep{davies97,jogee05}. If there is a low luminosity AGN in this galaxy, there is no influence of its emission in the circumnuclear ring \citep{mazzalay13}. The SMBH mass was estimated to be $log(M_{BH}/M_{\odot})= 7.577^{+0.052}_{−0.054}$  \citep{Berrier13}.

\textit{NGC 4689: }is a Milky-Way analogue classified non-barred SA(rs)bc \citep{RC3}. The galaxy has flocculent spiral arms \citep{jungwiert97}; however, \citet{arsenault89} says that it has a weak bar. Its nucleus is classified as starburst \citep{ho97}, confirmed by the non-
detection of X-ray emission, with a SMBH mass of $log(M_{BH}/M_{\odot})= 4.2 \pm 0.8$ \citep{graham19}. No molecular outflow was detected \citep{stuber21}. Molecular gas shows patchy morphology without clear structure (bar or spiral), with a regular velocity field \citep{sofue03,sofue03b}. Two pseudo rings were found, one located at the bar Inner Lindblad Resonance, 0'\!\!.57 $\times $0'\!\!.44, and other located at the Outer Lindblad Resonance, with 1'\!\!.86 $\times $1'\!\!.51  \citep{comeron14}.

\textit{NGC5248: } is an SAB(rs)bc \citep{RC3} galaxy with strong spiral arms \citep{elmegreen92}. Within our field of view, there is evidence of two rings, one inner ring of 650pc diameter around the bar and a double nuclear ring \citep{buta93,comeron10}. The bar fuels the  nuclear ring with diameter of 150 pc \citep{comeron10} with two molecular trailing spirals, which connect to other two arms \citep{jogee02}  \citep{buta93,maoz01,comeron10}. The nuclear rings have stellar population ages between 1 and 10 Myr \citep{elmegreen97}.  Its nucleus was classified as starburst \citep{ho97}
and Seyfert/LINER \citep{gadotti19} from the optical emission-line ratios, and harbours a SMBH with mass of $log(M_{BH}/M_{\odot})= 6.73^{-0.22}_{−0.20}$ \citep{she17}. There is no evidence of
molecular gas outflows \citep{stuber21}. \citet{haan09} studied the torques of the CO molecular gas and found that NGC 5248 shows a mixed pattern of gravitational torques that corresponds to the bar and spiral arms, comparable to our results.

\textit{NGC 5921: }is a SB(r)bc \citep{RC3} galaxy with a compact nucleus and a prominent bar that ends in an outer ring \citep{Eskridge02,fathi09}. Its nucleus is classified as composite \citep{ho97}, LINER and Seyfert-like \citep{veron-cetty86,arsenault89}, and non-active \citep{burtscher21}. This galaxy has an inner ring-like structure \citep{hernandez05,martinez-garcia12}. Its SMBH estimated mass is $log(M_{BH}/M_{\odot})=7.07^{-0.42}_{−7.07}$  \citep{vandenbosch16}.

\section{H$\alpha$ gas torques with MUSE data}

\begin{figure*}
\begin{center}

  \includegraphics[scale=0.21]{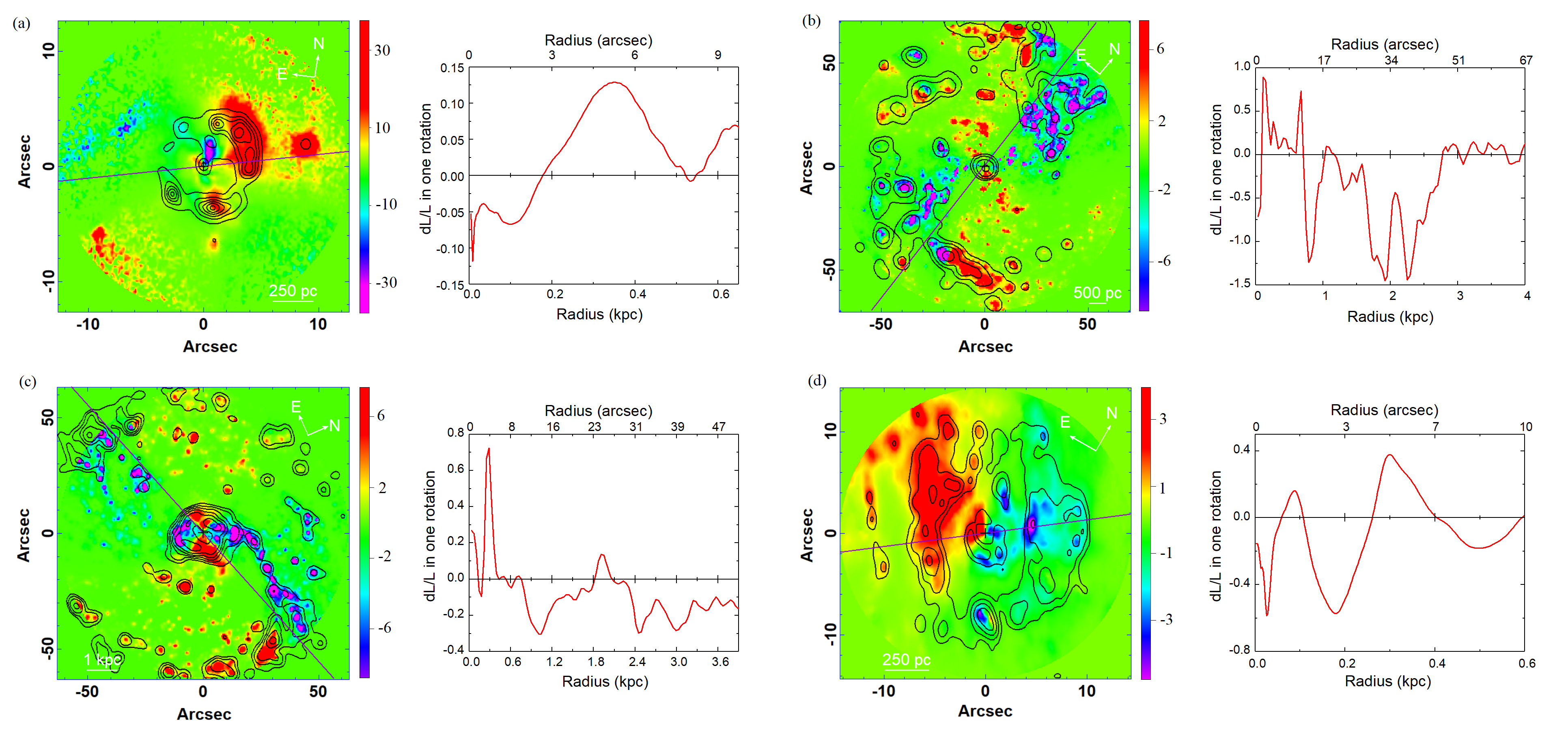}
  \caption{- Torque maps of H$\alpha$ emission line from MUSE data cube  of: (a)NGC 1300 with contours levels of  1 to 6 times 2.4 $\times 10^{-17}$ erg \AA $^{-1}$ cm$^{-2}$ s$^{-1}$, (b)NGC 4303 with contours levels of 1, 2.5, 6, and 14.3 times 1.75 $\times 10^{-17}$ erg \AA $^{-1}$ cm$^{-2}$ s$^{-1}$, (c) NGC 4321  with contours levels of 1, 2, 3.7, 7.1, 13.3 and 25.1 times 5.4 $\times 10^{-18}$ erg \AA $^{-1}$ cm$^{-2}$ s$^{-1}$, and (d) NGC 5248 with contour levels of 1, 1.8, 3.2 and 5.7 times 5.7 $\times 10^{-17}$ erg \AA $^{-1}$ cm$^{-2}$ s$^{-1}$.} \label{torquemaps_N5248b}

\end{center}
\end{figure*}

For the sample of 10 galaxies of this work, only four galaxies (NGC 1300, NGC 4303, NGC 4321 and NGC 5248) had Multi Unit Spectroscopic Explorer (MUSE) data cubes available. The H$\alpha$ emission line can trace the star forming regions and interstellar medium in the less dense regions, giving a more complete image of the structures of the galaxy. In contrast, the molecular gas emission is more compact, giving a overall idea of the structures punctually. In order to see the distribution of torques in general in the galaxies and compare them with the obtained molecular gas torques, we made images of H$\alpha$ emission line from MUSE data cubes, due to the good spatial resolution, and FOV, to compute the torques on this gas. The data were taken from the ESO public archive. The information about the observations of each galaxy is shown in table \ref{obsinfo_MUSE}. 

The H$\alpha$ emission-line images were taken from the data cubes by summing the spectral interval of the emission line and by subtracting the nearby continuum equivalent to the same interval. After that, the torques were computed associated to the red images based on the size of the FOV. In the case of NGC 1300, we used HST (cyan square in Fig.~\ref{galaxiesimage} or Fig.~\ref{NGC1300_m0_m1}b), in the case of NGC 4303, the Legacy image was used (red square in Fig.~\ref{galaxiesimage} or Fig.~\ref{NGC4303_m0_m1}b), for NGC 4321, we also used the Legacy image (red square in Fig.~\ref{galaxiesimage} or Fig.~\ref{NGC4321_m0_m1}b), and for NCG 5248 we used the HST image (cyan square in Fig.~\ref{galaxiesimage} or Fig.~\ref{NGC5248_m0_m1}b). The FOV of the MUSE data was cut to be the same size of the associated FOVs. We then computed the torques analogously to what was done with the ALMA data. 

\begin{table}
\caption{- Observational information of MUSE data from NGC 1300, NGC 4303, NGC 4321, and NGC 5248.}\label{obsinfo_MUSE}
\resizebox{\columnwidth}{!}{%
\begin{tabular}{ccccc}
\hline
Galaxy   & Program ID    & P. I.          & Observation Date & Exposition time (s) \\ \hline
NGC 1300 & 097.B-0640(A) & Gadotti, D.    & October 2nd 2016 & 2880             \\
NGC 4303 & 1100.B-0651   & Schinnerer, E. & May 10th 2019    & 2580            \\
NGC 4321 & 1100.B-0651   & Schinnerer, E. & April 28th 2019  & 2580           \\
NGC 5248 & 097.B-0640(A) & Gadotti, D.    & April 4th 2016   & 3840            \\ \hline
\end{tabular}%
}
\end{table}

\end{appendix}


\end{document}